\documentclass[prd,a4paper,reprint,superscriptaddress,preprintnumbers,nofootinbib]{revtex4-2}

\usepackage{mymacros}

\definecolor{change}{rgb}{0.0, 0.0, 1.0} 

\begin{document}

\title{A quantum information perspective on meson melting}

\author{Mari~Carmen~Ba\~nuls} 
\email{mari.banuls@mpq.mpg.de}
\affiliation{Max Planck Institute of Quantum Optics, 85748 Garching bei M{\"u}nchen, Germany}
\affiliation{Munich Center for Quantum Science and Technology (MCQST), 80799 M{\"u}nchen, Germany}

\author{Michal P.\ Heller} 
\email{michal.p.heller@ugent.be}
\affiliation{Department of Physics and Astronomy, Ghent University, 9000 Ghent, Belgium}

\author{Karl Jansen} \email{karl.jansen@desy.de}
\affiliation{Deutsches Elektronen-Synchrotron (DESY), Platanenallee 6, 15738 Zeuthen, Germany}

\author{Johannes Knaute} \email{johannes.knaute@mail.huji.ac.il}
\affiliation{Racah Institute of Physics, The Hebrew University of Jerusalem, Jerusalem 91904, Givat Ram, Israel}

\author{Viktor Svensson} 
\email{viktor.svensson@ftf.lth.se}
\affiliation{Division of Solid State Physics and NanoLund, Lund University, S-221 00 Lund, Sweden}

\begin{abstract}
We propose to use quantum information notions to characterize thermally induced melting of nonperturbative bound states at high temperatures. We apply tensor networks to investigate this idea in static and dynamical settings within the Ising quantum field theory, where bound states are confined fermion pairs -- mesons. 
An equilibrium signature of meson melting is identified in the temperature dependence of the thermal-state second R\'enyi entropy, which varies from exponential to power-law scaling. Out of equilibrium, we identify as the relevant signature the transition from an oscillatory to a linear growing behavior of reflected entropy after a thermal quench. These analyses apply more broadly, which brings new ways of describing in-medium meson phenomena in quantum many-body and high-energy physics.
\end{abstract}
\maketitle

\tableofcontents

\section{Introduction and motivation}

Emergent phenomena of quantum field theories (QFTs) under extreme conditions can pose significant challenges for their theoretical description both in condensed matter and particle physics \cite{coleman_2015,Greiner_2007,Busza:2018rrf,Berges:2020fwq}. Many problems of this type are motivated by quantum chromodynamics (QCD), as is the focus of this article: the melting of mesons~\cite{Rothkopf:2019ipj,Friman:2011zz}. The phenomenon of meson melting, i.e. the breakdown of bound states into their elementary constituents at high enough temperatures, is of fundamental interest in QCD, where it is relevant for the physics of the early universe and for the understanding of nuclear collisions. Whereas a full explanation of the process from first principles is not available, the current phenomenological description understands the meson melting as a dynamical process in which out of equilibrium mesons interact with a thermal background~\cite{Rothkopf:2019ipj}. This phenomenology is not exclusive of QCD, though, but can be explored in other models that exhibit mesons in their spectrum.

In this article, we propose a new approach to the meson melting phenomenon through the study of entanglement measures. As a first step in this direction, we consider the ($1+1$)-dimensional Ising QFT, which contains both integrable and nonintegrable parameter regimes exhibiting meson excitations. In particular, we study static and time-dependent thermal states in different temperature regimes in the thermodynamic and continuum limit of the quantum Ising model near its critical point by analyzing several entropic quantities. For this purpose we employ numerical tensor network (TN) simulations, a tool originating from quantum information that directly gives access to both entropic quantities and entanglement measures \cite{Verstraete08,SCHOLLWOCK201196,Orus:2013kga,Okunishi2022review,Banuls2022review}. The dynamical setup is generated by a quantum quench -- a common method, which creates an out of equilibrium state through a (instantaneous) parameter change in the underlying Hamiltonian. This protocol can be seen as a theoretical method by itself, since it can induce entirely new dynamical features to the system. We use it to mimic a dynamical situation as in the modern QCD viewpoint on meson melting. 
Our study of entanglement measures in the context of meson melting can be seen complementary to the works~\cite{Kharzeev:2014pha,Hashimoto:2014fha,Iatrakis:2015sua}, in which quarkonium suppression and dissociation have been described as an entropic self-destruction.
Notice also that, recently, entanglement measures have been used in a similar vein to characterize ($1+1$)-dimensional meson scattering events~\cite{Rigobello:2021fxw}. 

Our TN approach is also closely related to the field of quantum simulations \cite{Hauke2011d,Cirac2012,Alexeev:2020xrq,Fraxanet:2022wgf}, in which quantum hardware is utilized to study many-body problems via digital or analog implementations. Recently, there has been a considerable amount of attention to the potential use of these technologies for studies of meson physics, both in experimental measurements \cite{Tan:2019kya,Vovrosh_2021,Vovrosh:2021ocf,Mildenberger:2022jqr} as well as in theoretical proposals \cite{Verdel:2019chj,Surace:2020ycc,Karpov:2020pqe,Knaute:2021xna,Vovrosh:2022bpj}. With such arising quantum technologies offering a significant long-term promise for studying fundamental physics problems, it is a timely problem to frame the melting of bound states using their native language of quantum information.

\subsection*{Outline of the article}

In describing the effect of melting mesons from a quantum information perspective, we are crossing several fields in physics. To make our work accessible for a broad audience, we aim for a self-contained presentation and organize this article as follows.

Sections~\ref{sec:prelim} and \ref{sec:methods} are devoted to a short review of physical and conceptual background material relevant for our studies. In particular, section~\ref{sec:prelim} contains a discussion of the meson melting phenomenology in QCD, which primarily motivates our work (section~\ref{sec:prelim_pheno}). We then introduce in section~\ref{sec:prelim_Ising} the model we analyze, the 1+1$D$ Ising QFT, which emerges in the continuum limit of the quantum Ising model. In section~\ref{sec:methods} we describe relevant concepts of entanglement properties in quantum many-body systems, which underlie TN ans\"atze and numerical algorithms for our simulations. This section also contains a discussion of entropic quantities and quantum quenches.

The results of our analyses are discussed in sections~\ref{sec:Renyi} and \ref{sec:quench}. In section~\ref{sec:Renyi} we present simulation results for different scaling behaviors of the R\'enyi entropy density in thermal equilibrium states. In section~\ref{sec:quench} we analyze entanglement entropies after thermal quantum quenches at different effective temperatures. In both cases, we consider mesonic parameter regimes of the Ising QFT and find signatures that indicate the melting of meson states at high temperatures.

We devote several appendices to further background information and computational details. In Appendix~\ref{app:review} insights from methods in QCD and holography are discussed, while in Appendix~\ref{app:TNS} more details on TNs are presented. Details on a transfer operator method and signal analysis techniques can be found in Appendices~\ref{app:transferop} and \ref{app:prony}, respectively.

\section{Preliminaries}
\label{sec:prelim}

\subsection{Phenomenology of the meson melting process in QCD}
\label{sec:prelim_pheno}

Spectral functions are the main quantities studied within QFT to describe the melting of mesons. In this section we provide an introduction to the known phenomenology of this effect based on the recent and comprehensive review \cite{Rothkopf:2019ipj}. For more detailed discussions and further background information we refer to Appendix~\ref{app:review_methods}. 

In thermal equilibrium the spectral function $\rho$ is given by the negative imaginary part of the retarded correlator,
\begin{equation} \label{eq:rhoImDR}
    \rho(\vec p,\omega) = -\operatorname{Im}[D_R(\vec p,\omega)] .
\end{equation}
Here, the retarded correlator $D_R$ is defined w.r.t.\ a \textit{meson operator} $M(\vec x,t)$ as the thermal expectation value $D_R(\vec{x},t,\vec{x}_0,t_0) \equiv \theta(t-t_0) \Tr\{\rho_\beta [M(\vec x,t), M^\dagger(\vec x_0,t_0)] \}$, where $\rho_\beta$ is the thermal density operator at inverse temperature $\beta$ and $\theta(t)$ is the Heaviside function. In a QCD context, $\rho$ is typically considered in Fourier space via a Wigner transformation, in which relative frequency and momentum coordinates $(\vec p,\omega)$ are used as in \eqref{eq:rhoImDR}.
The spectral function in this form probes the particle content of the physical system. Meson bound states show up as peaks at frequency values given by their masses.  

Within QCD, there are several established methods to calculate spectral functions of quark bound states.\,\footnote{We refer to Appendix~\ref{app:review_methods} for a short overview and discussion of them.} They form the basis for our current understanding, which was initiated by the influential work \cite{Matsui:1986dk} on suppression and melting of mesons. Starting from an analogy with Debye screening in an electromagnetic plasma, the authors argue that the quark-gluon plasma created in a nuclear collision weakens the binding energies of meson bound states in the thermal environment, causing a suppression of meson detection rates.
Building up on that, the work~\cite{Karsch:2005nk} put forward the picture of a sequential process. In this static scenario, an in-medium Hamiltonian describes mesons as its eigenstates. The potential term in the Hamiltonian gets weaker for increasing temperatures, such that bound states transform into scattering states at a unique \textit{melting temperature}, where a sharp transition takes place. Weakly bound meson states melt first, while strongly bound ones survive longer in the thermal medium.

The work \cite{Laine:2006ns} opened a new way of thinking about this process by showing that the quark potential \eqref{eq:V_s} is complex. This implies that a meson bound state is a dynamical system, which interacts constantly with its thermal environment by scattering processes. In this modern picture, the melting process is best understood from the behavior of spectral functions in dependence of the temperature. With increasing temperature, the meson peaks move towards smaller frequency values; their magnitude is decreasing while their thermal width is increasing. Peaks at larger frequencies, i.e.\ lower binding energies, melt first and their peak structure dissolves completely into a continuum. The combination of thermal broadening and decreasing of the binding between the quarks means that mesons transition into other (unknown) states. The dynamical origin of this process is not known from this description. As a consequence, the definition of a melting temperature in this time-dependent, sequential scenario is not uniquely determinable any more and a less important concept. The overall intuitive picture drawn in \cite{Rothkopf:2019ipj} is that the thermal environment acts as a sieve that filters out weakly bound meson states, which are exposed to intensifying scattering events as the temperature is raised. Apart from direct approaches in QCD, also holographic models corroborated this phenomenological picture of a sequential melting process. Some insights from these methods are reviewed in Appendix~\ref{app:review_holo}.

From these elaborations, it becomes obvious that only a fully real-time treatment can describe all dynamical processes contributing to the melting of meson bound states. Recently, open quantum system approaches to this problem provided new insights in this direction. They treat the interactions between a system (quark-antiquark pair) and its environment (thermal medium) in a real-time Schr\"odinger formalism. Interestingly, it could link the melting process to the quantum mechanical effect of decoherence (see e.g.\ \cite{Kajimoto:2017rel,DeBoni:2017ocl} and further references in \cite{Rothkopf:2019ipj}).

\subsection{The 1+1$D$ Ising QFT}
\label{sec:prelim_Ising}

The Ising spin model is defined by the Hamiltonian
\begin{equation}  \label{eq:H_NN}
H = - J \,\left ( \sum_{j=1}^{N-1} \sigma^z_j \sigma^z_{j+1} + h \sum_{j=1}^N  \sigma^x_j + g \sum_{j=1}^N \sigma^z_j \right) ,
\end{equation}
where $\sigma_j^{\alpha}$ is the corresponding Pauli matrix at position $j$, and we have considered an open chain of $N$ sites. Throughout our numerical calculations, we use $J \equiv 1$. This sets an overall energy scale associated with the lattice. Furthermore, $h$ and $g$ are, respectively, the transverse and longitudinal fields.

In the pure transverse field Ising model ($g=0$), it is well known that a quantum phase transition takes place at the critical point $J=h=1$ from a ferromagnetic (ordered) phase ($h<1$) towards a paramagnetic (disordered) phase ($h>1$). In the thermodynamic (infinite system size $N \to \infty$) and continuum limit (vanishing lattice spacing $a \equiv 2/J \to 0$), the infrared (IR) regime of the Hamiltonian \eqref{eq:H_NN} is effectively described by the Majorana fermion QFT \cite{Rakovszky:2016ugs}
\small
\begin{equation} \label{eq:H_IsingQFT}
H_\text{IR} = \int_{-\infty}^\infty \mathrm dx \,  \left\{ \frac{i}{4\pi} \left( \psi\partial_x\psi - \bar\psi\partial_x\bar\psi \right) - \frac{i M_h}{2 \pi}\bar\psi\psi + {\cal C} M_{g}^{15/8} \, \sigma \right\} .
\end{equation}
\normalsize
The free fermion mass $M_h$ is related to the transverse lattice parameter $h$ through the relation $M_{h} \equiv 2 J |1-h|$, while the longitudinal mass scale is given by $M_{g} \equiv \, {\cal D} J \, |g|^{8/15}$. Here, ${\cal C} \approx 0.062$ and ${\cal D} \approx 5.416$ are numerical constants~\cite{Rakovszky:2016ugs, Hodsagi:2018sul}. 

As discussed in detail in \cite{Rakovszky:2016ugs}, the QFT defined by the Hamiltonian \eqref{eq:H_IsingQFT} appears in the scaling limit $M_h/J \to 0$ for fixed ratio $M_h/M_g$. It comprises the following important classes in the vicinity of the critical point. First, if $M_h = M_g = 0$ (i.e.\ at the critical point $\{h=1, g=0\}$), the Hamiltonian \eqref{eq:H_IsingQFT} represents the Ising conformal field theory (CFT), which is a free Majorana fermion CFT with central charge $c = \frac{1}{2}$. There are two scalar primary Hermitian operators, 
\begin{equation}
\label{eq.epsdef}
\epsilon \equiv i\bar\psi\psi \sim \sigma_j^x ,   
\end{equation}
with scaling dimension $\Delta_\epsilon=1$ and $\sigma \sim \sigma_j^z$ with $\Delta_\sigma = \frac{1}{8}$. For $M_h \ne 0$, $M_g=0$, the continuum limit of the transverse Ising model in both the ferromagnetic and paramagnetic phase is integrable in terms of a massive free fermion QFT, characterized by the fermion mass $M_h$. Elementary excitations in the ferromagnetic phase can be interpreted as domain walls. As soon as a longitudinal field is turned on ($g \ne 0$), these are non-perturbatively confined \cite{McCoy:1978ta}. The two remaining important regimes therefore give rise to mesonic excitations with the following properties:
\begin{itemize}
    \item $M_h=0$, $M_g \ne 0$: This regime describes the integrable and interacting E$_8$ QFT of Zamolodchikov \cite{Zamolodchikov:1989fp}, which is mathematically captured by the exceptional simple Lie algebra of rank 8. It contains 8 stable mesons as nonperturbative fermionic bound states in the spectrum. Their masses $M_n$ are known and can be expressed as analytical ratios in terms of the lightest mass $M_1 \equiv M_g$.
    \item $M_h \ne 0$, $M_g \ne 0$: This parameter range with both $\epsilon$ and $\sigma$ perturbations turned on represents an interacting nonintegrable QFT with stable and unstable meson bound states~\cite{Zamolodchikov:1989fp,Delfino2005Jul,Fonseca:2006au,Zamolodchikov:2013ama}.
\end{itemize}
We are primarily interested in studying the latter nonintegrable regime at non-zero temperature, in which TN simulations have the most predictive power since they provide an ab initio way of capturing the underlying meson physics. Despite the simplicity of the original lattice model \eqref{eq:H_NN}, we therefore can study a highly nontrivial class of QFTs in \eqref{eq:H_IsingQFT}, which we denote as the ($1+1$)-dimensional Ising QFT in the following discussion.

\section{Methods and concepts}
\label{sec:methods}

\subsection{Tensor network simulations}

We perform numerical simulations using matrix product states (MPS) and matrix product operators (MPO)~\cite{Verstraete08,SCHOLLWOCK201196,Cirac2009,Orus:2013kga}. These are TN ans\"atze that provide efficient representations of the state and operators of the quantum many-body system (see Appendix~\ref{app:TNS} for details). More concretely, a MPS parametrizes a quantum many-body state of a lattice system by a set of tensors (one per site) of dimensions $d\times \chi \times \chi$, where $d$ is the physical dimension of the degree of freedom associated to each lattice site, and $\chi$, called bond dimension, determines the size of the variational family, and bounds the entanglement entropy of the MPS. In the broadest sense, a MPO is simply a MPS for the space of operators, where the local dimension is $d^2$ instead of $d$.

The MPS/MPO ansatz allows us to work directly in the infinite spatial volume (thermodynamic limit), by assuming translational invariance and studying a unit cell. For a nearest-neighbor model like~\eqref{eq:H_NN}, we use a two-site unit cell and apply the infinite time-evolving block decimation (iTEBD) algorithm \cite{Vidal:2006ofj} to simulate both imaginary and real-time evolution. The former allows us to efficiently approximate thermal states as an MPO \cite{Hastings2006a,molnar2015thermal}, while the latter is employed in the subsequent quench studies. The iTEBD algorithm is based on a Suzuki-Trotter decomposition (see~\cite{Hatano:2005gh} and references therein) of the time evolution operator into discrete time steps and a sequential application of two-body operators acting on the MPS or MPO ansatz. The resulting bond dimension increases, in general, under this operation, and in practice it is necessary to truncate it. This is one source of error in the numerical algorithms. While it also limits the reachable time scales under real-time evolution, this will prove not to be a significant obstacle for the problem at hand.

\subsection{Entropies}
\label{subsec:entropies}

In our work several entropic quantities play a pivotal role in unraveling the meson melting. In particular, we make use of the von Neumann entropy and of the more general R\'enyi entropies. The former is defined as
\begin{equation} \label{eq:EE}
    S(\rho) = -\Tr \left[\rho \ln \rho \right] ,
\end{equation}
where $\rho$ is the density matrix describing the state of the system. The R\'enyi entropies~\footnote{Observe that in the limit $\alpha \rightarrow 1^+$, $S_\alpha(\rho_A)$ reduces to $S(\rho_A)$. In the continuum or in presence of gauge fields, the definition of reduced density matrices can lead to mathematical subtleties, which, however, will not be relevant for our subsequent spin chain calculations.}  are defined as
\begin{equation} \label{eq:RenyiE}
    S_\alpha(\rho) = \frac{1}{1-\alpha} \ln(\Tr\rho^\alpha).
\end{equation}
In the case of pure states, when we consider a bipartition in subsystems A and B, the corresponding entropies for the reduced density matrix $\rho_A \equiv \Tr_B \rho$ of a subsystem measure the entanglement with respect to this bipartition, and the von Neumann entropy~\eqref{eq:EE} $S(\rho_A)=S(\rho_B)$ is then referred to as the entanglement entropy. 

In the case of pure states given by MPS, these entropies can be efficiently calculated, because it is easy to recover the Schmidt decomposition across each cut between a pair of sites, which in turn provides the eigenvalues of the corresponding reduced density matrix for half the chain.
In the case of mixed states given by MPO, like the ones representing thermal states, these quantities cannot be efficiently recovered, in general.  
We can however compute the 2-R\'enyi entropy (or another low integer order) of the whole state.

Another option utilized by us is to consider a purification of the mixed state, i.e.\ embedding a given mixed state as a reduced density matrix of a pure state from an enlarged Hilbert space. A canonical purification of a mixed state, demonstrated here on an example of a thermal density matrix, is given by the thermofield double state 
\begin{equation} \label{eq:puri}
\ket{\Psi_{\beta}} \propto \e^{-\beta H/2} \ket{\Phi},
\end{equation}
where $|\Phi\rangle$ is a maximally entangled state in a doubled system. Such a canonical purification can be constructed also for a general mixed state. For the canonical purification there is a natural identification of spatial subsystems in the original system and corresponding subsystems in the copy. This leads to the notion of the \textit{reflected entropy}, which is the entropy associated with tracing out the same spatial subsystems in both Hilbert space factors in the canonical purification. When the purification is given as a MPS, one can efficiently compute the reflected entropy using the MPS/MPO techniques, which will play an important role in the rest of the paper. The reflected entropy was originally introduced in \cite{Dutta:2019gen} in a holographic setting as a proposed dual to twice the area of the entanglement wedge cross section. Because of this, it has received a considerable amount of attention in the context of the AdS/CFT correspondence and QFTs, see e.g.\ \cite{Jeong:2019xdr,Bao:2019zqc,Chu:2019etd,Bueno:2020vnx,Li:2020ceg,Bueno:2020fle,Camargo:2021aiq,Hayden:2021gno}. In particular, the works \cite{Kudler-Flam:2020url,Moosa:2020vcs,Kudler-Flam:2020xqu} discussed it also for CFT quenches, whereas \cite{Jain:2020rbb} analyzed a variety of related mixed state entanglement measures as a holographic probe of confinement.

The purification of a mixed state is not unique, since any unitary transformation on the ancillary copy leaves the state of the system unchanged. The entanglement entropy is in general not invariant under such transformation. One can define a version of the reflected entropy, called the entanglement of purification~\cite{Terhal:2002}, as the minimal one over all possible purifications. The reflected entropy provides an upper bound to such an optimal quantity, which in general is hard to compute. Note that the area of the entanglement wedge cross section is conjectured to correspond to the entanglement of purification~\cite{Takayanagi:2017knl}. If this conjecture is correct, it implies that in holographic states optimization allows lowering the reflected entropy by a factor of 2 by applying a unitary on a purifying system.

In the present work, we employ the TN ansatz to extract the discussed entropic quantities for the Ising QFT. Before closing this section, we want to point out that the authors of \cite{Murciano:2021huy,Murciano:2021dga} recently developed a simulation scheme to calculate (generalized) R\'enyi entropies for bosonic systems and 1+1$D$ CFTs. In a parallel vein, \cite{Emonts:2022vim} calculated entanglement measures in ($1+1$)-dimensional theories using a Hamiltonian truncation approach~\cite{Yurov:1991my}, which provides an alternative computational ansatz.

\subsection{Quantum quenches and quasiparticle model}

We now turn to the concept of quantum quenches to study a dynamical situation in the context of meson melting. A quantum quench means a rapid change of a Hamiltonian parameter governing the time evolution of a physical system. Initiated primarily by the\linebreak works~\cite{Calabrese:2005in,Calabrese:2006rx,Calabrese:2007rg}, quantum quenches became a theoretical framework to probe and characterize dynamical properties of quantum many-body systems~\cite{mitra2018quench}. The modification of the underlying Hamiltonian during time evolution kicks the system out of equilibrium. This drastic change can induce new processes and effects through dynamical interactions. 

Our studies are motivated by the seminal work~\cite{kormos2017real}. Therein, the authors analyzed the entanglement growth for quantum quenches in the Ising model. In more detail, iTEBD methods were used to prepare the ground state $\ket{\psi_0}$ of some Hamiltonian $H_0$, which then was evolved with a Hamiltonian $H_1$ in the ferromagnetic phase in presence of meson states, i.e.\ at finite longitudinal field values. It was observed that the existence of mesons causes oscillations in the entanglement entropy at frequencies given by their masses and mass differences. If the initial state is in the paramagnetic phase (i.e.\ a quench across the phase boundary), the associated entanglement growth persists mostly forever, whereas it is always bounded if $\ket{\psi_0}$ is located in the ferromagnetic phase and mesons are produced at rest. While the original work \cite{kormos2017real} considered a semiclassical parameter regime away from the quantum critical point, the recent paper \cite{Castro-Alvaredo:2020mzq} observed such \textit{entanglement oscillations} also in the E$_8$ QFT regime of the Ising model. Similar observations were made for gauge theories in \cite{Magnifico:2019kyj,Chanda:2019fiu,Lerose:2019jrs} as well as for quasiparticles in paramagnetic quenches~\cite{Collura_2018,Pomponio_2019}. Further aspects of entanglement dynamics and thermalization behavior in spin chain models are discussed, e.g., in \cite{Leviatan:2017vur,Lencses:2020jjt,Birnkammer:2022giy}.

Many properties of the entanglement production and spreading after quantum quenches can be understood with a \textit{quasiparticle model}. Originally developed in \cite{Calabrese:2005in} for the Ising model, it was later found to hold also for generic integrable systems (in absence of mesons) \cite{Alba_2017,Alba:2017lvc,DeLuca:2019vns} and thermofield double states \cite{Chapman:2018hou}. A generalization of the model was provided in~\cite{Bastianello_2020}. 
In essence, the model describes the initial state as a source of independent entangled quasiparticle pairs, which are created at any given point and move after the quench with opposite momentum and velocity through the quantum many-body system and therefore spread quantum correlations. In lattice systems, the maximum speed of propagation is limited by the Lieb--Robinson bound \cite{Lieb:1972wy}. As a consequence, the entropy of a finite-size subregion will grow linearly in time as pairs spread with one of their members crossing the boundary of the region. Eventually, the entanglement saturates when both members of all pairs produced inside the subsystem have left it. This picture is consistent with general entanglement scaling laws in the time evolution found in \cite{Eisert:2006zz}. For the special case of a semi-infinite spatial bipartition this implies an indefinite entanglement growth. However, the boundedness of entanglement production for the meson case can also be explained within the quasiparticle picture~\cite{kormos2017real,Scopa:2021gcx}. Namely, the existence of a confining potential, due to the presence of meson states, induces that quasiparticles bounce back as they get separated from their partners, causing the oscillatory and bounded behavior for mesons produced at rest. 

In this paper, we consider quenches in which the system is initially prepared in a thermal equilibrium state, and then is evolved with a different Hamiltonian. Such finite temperature quenches are much less explored than those starting from a ground state, see \cite{Sotiriadis_2009,Zhang_2011,Bacsi_2013,Collura_2014,Granet:2020xuj,kohn2021quenching} for selected results. In this scenario, we analyze the time evolution of different entropic quantities to detect the melting of mesons at high temperatures. For nonintegrable parameter regimes, in which we are primarily interested here, it was shown in \cite{Scopa:2021gcx} that quantitative quasiparticle model interpretations are applicable only at low energies (or temperatures), when mesons are dilute and one can focus on the two-fermion problem. Since we enter physical situations beyond this case, we will make use of quasiparticle interpretations in our subsequent meson analyses only in a qualitative fashion, or compare to the free fermion regime.

\section{Results from thermal equilibrium}
\label{sec:Renyi}

\begin{figure*}[t]
    \centering
   \includegraphics[width=0.49\textwidth]{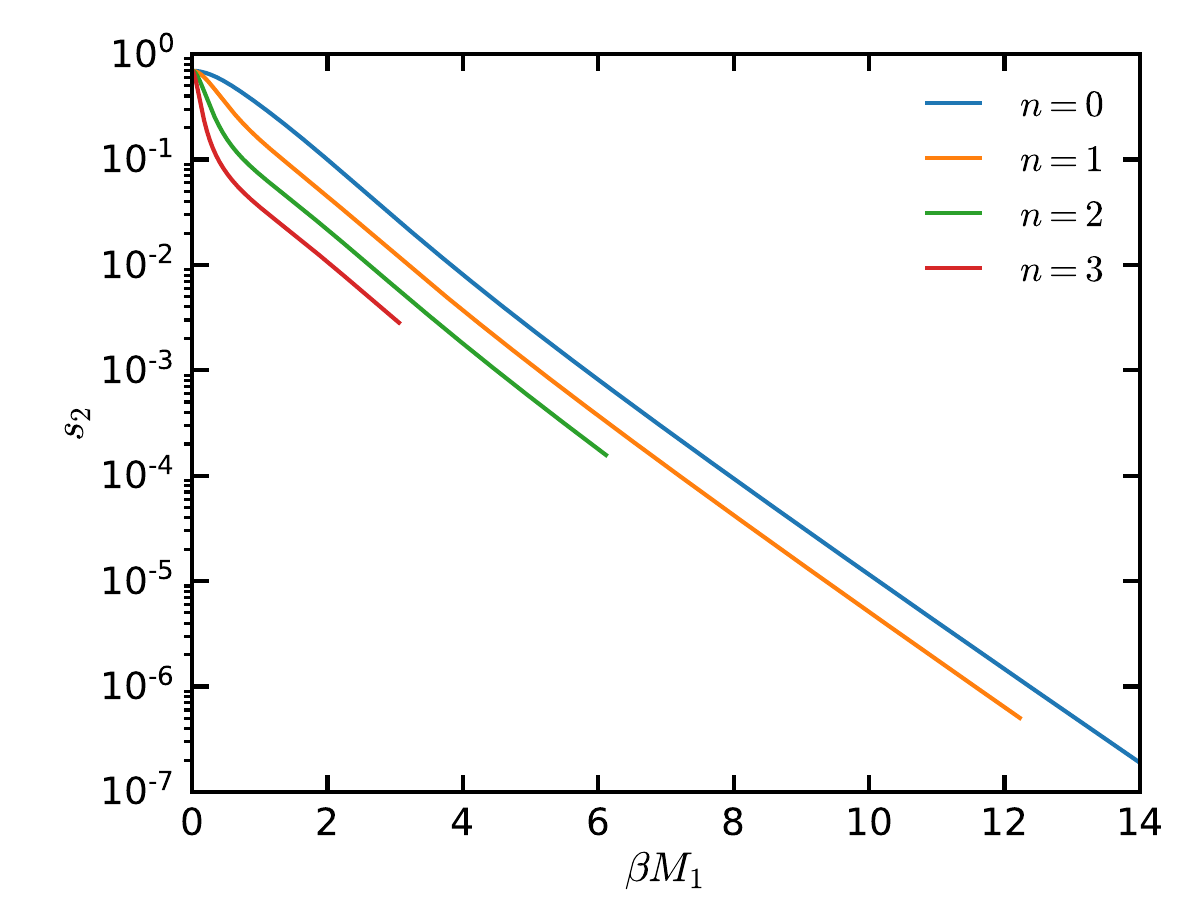} 
   \includegraphics[width=0.5\textwidth]{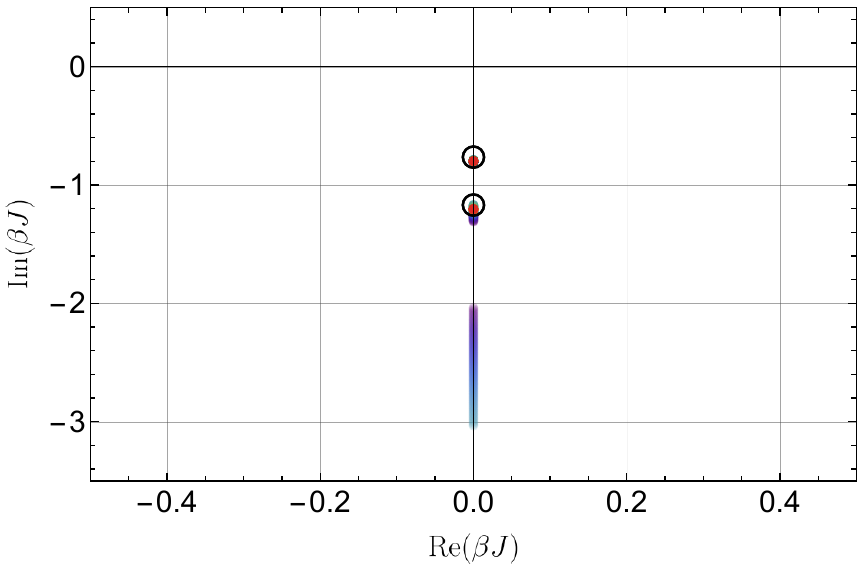}
    \caption{Left: Temperature dependence of the R\'enyi entropy density $s_2$. The curves are calculated using iMPO simulation at constant ratio $M_h/M_g \approx 0.09$ in the nonintegrable ferromagnetic phase using the parametrization \eqref{eq:Mhg_params}.
    Right: Prony result of the $n=1$ curve of $s_2(\beta J)$ in the complexified $\beta J$ plane. 
    The plots demonstrate that the low temperature behavior of $s_2$ is dominated by an exponential decay with frequencies matching the meson masses $M_1/J$ and $M_2/J$ (extracted from \cite{Fonseca:2006au}), shown as black circles in the right panel.}
    \label{fig:s2_lowT_scaling}
\end{figure*}

\begin{table}[t]
    \centering
\caption{\label{tab:Renyi_scaling}High-temperature scaling exponents $p$ in the functional power-law ansatz \eqref{eq:power_law_scaling} for several quantities $f$ and the parametrization \eqref{eq:Mhg_params}. The numerical values are obtained from a fit to the iMPO simulations and are in good agreement with the analytical CFT expectation in the scaling limit.}    
     \begin{tabular}{l | *{4}{c} | c} \hline
$f$       & $n=0$ & $n=1$ & $n=2$ & $n=3$ & exact \\ \hline 
$s_2$     & 0.926 & 1.073 & 1.106 & 1.078 & 1     \\ 
$E$       & 1.171 & 1.645 & 1.933 & 2.022 & 2     \\ 
$\partial E/\partial\beta$ & 1.176 & 2.027 & 2.709 & 2.955 & 3 \\ \hline
\end{tabular}
\end{table}

\begin{figure}[t]
    \centering
   \includegraphics[width=0.49\textwidth]{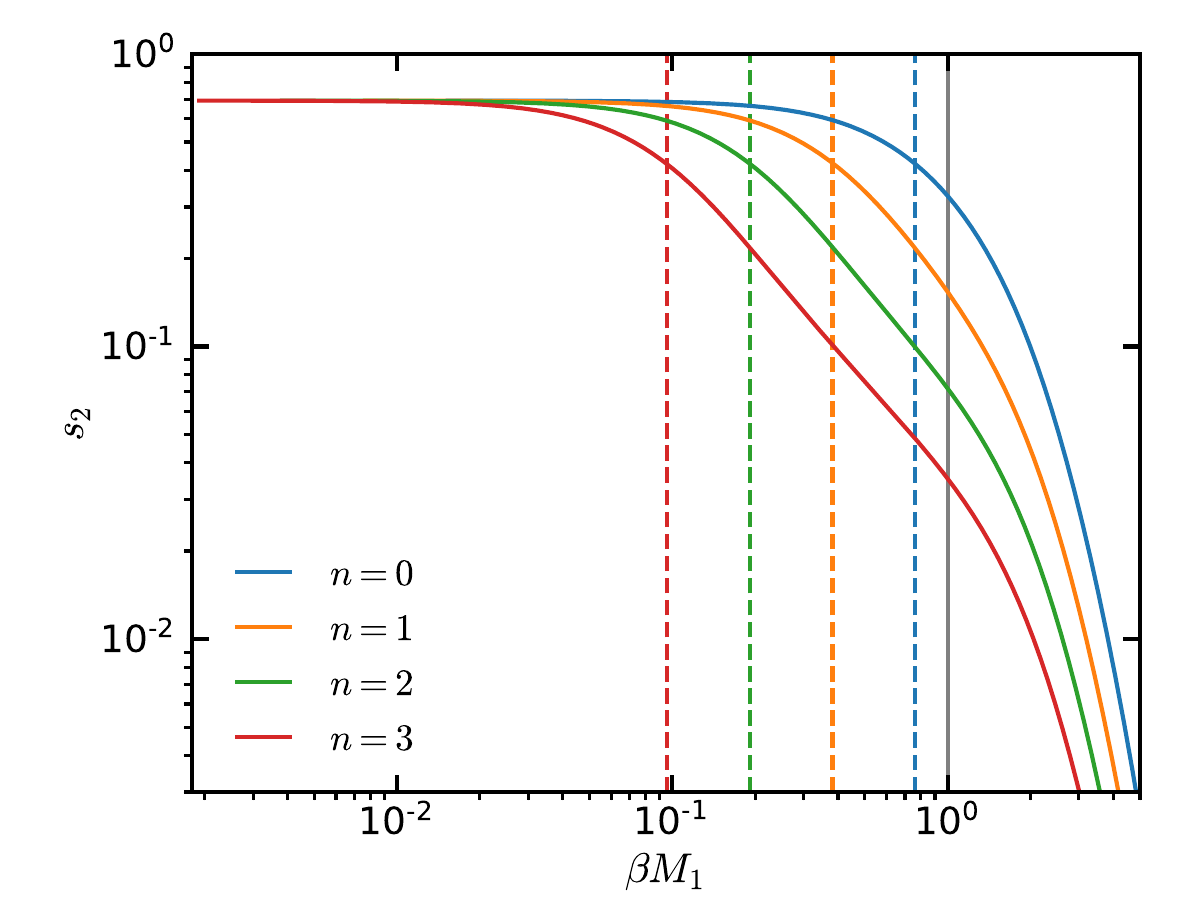} 
   \includegraphics[width=0.49\textwidth]{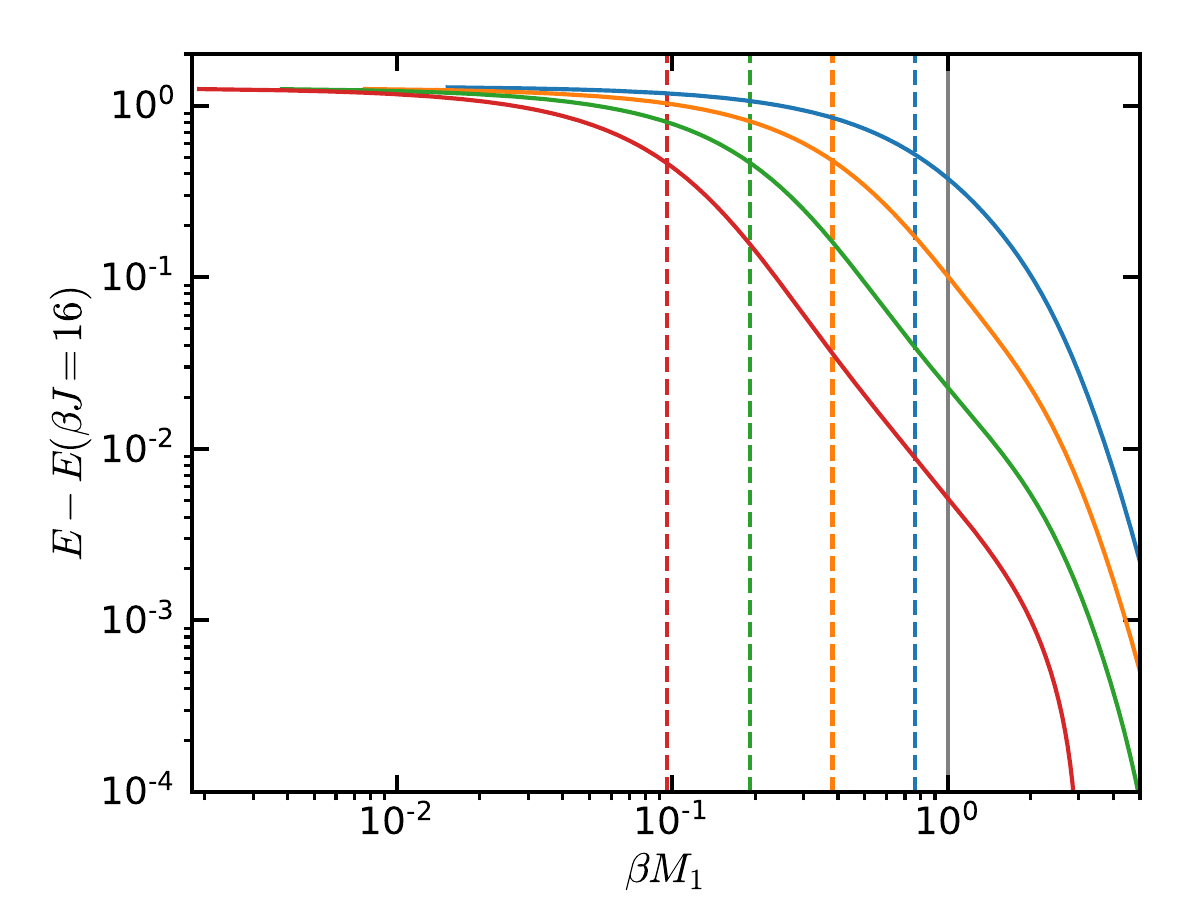} 
    \caption{High-temperature scaling of the second R\'enyi entropy density $s_2$ (top) and thermal energy density $E$ (bottom) in the nonintegrable ferromagnetic phase at $M_h/M_g \approx 0.09$ for the parametrization \eqref{eq:Mhg_params}. The gray line at $\beta M_1=1$ indicates the high-temperature threshold. Dashed lines indicate the lattice scale for each curve. When taking the scaling limit (increasing $n$) the results indicate the emergence of a power-law behavior matching the CFT expectation, cf.\ Tab.\,\ref{tab:Renyi_scaling}.}
    \label{fig:s2_highT_scaling}
\end{figure}

In section~\ref{sec:prelim_pheno} we have discussed that in a QCD context meson melting is indicated by a thermal broadening and movement of peaks in the in-medium spectral function. For the Ising QFT, our previous work \cite{Banuls:2019qrq} studied the retarded thermal correlation function w.r.t.\ the $\epsilon$ operator~\eqref{eq.epsdef}, which is the closest analog in this system to the simplest meson operator in QCD. We did observe sharp meson peaks, whose residues decrease with the temperature. Further, we observed that growing temperature causes the appearance of singularities associated to meson mass differences, which we interpreted as in-medium effects (for detailed discussions of both effects see also \cite{Knaute:2022ndb}). However, over the range of the temperatures explored we did not see definite signatures of the meson broadening. This, of course, does not mean that the mesons do not melt, but rather that the chosen observable in conjunction with the range of parameters and the methods adopted are not suitable to observe such an effect. In particular, our previous studies required long time evolutions to resolve the peaks associated with the lowest lying mesons, which became prohibitively costly for TN methods due to the required size of the bond dimension at large temperatures.

For this reason in the present work we shift our viewpoint to the study of entropies in the ferromagnetic regime of the Ising QFT. These quantities are sensitive to classical and quantum correlations in the many-body system, and can provide a complementary probe of meson features at high temperatures. In this section, we consider a static setting, namely thermal equilibrium states of the Ising model \eqref{eq:H_NN} at inverse temperature $\beta$, and study their second R\'enyi entropy per site, defined as
\begin{equation} \label{eq:s2}
    s_2 = -\frac{1}{N}\ln\frac{\Tr\rho_\beta^2}{(\Tr\rho_\beta)^2} 
\end{equation} 
for the (unnormalized) thermal state $\rho_\beta = \exp[-\beta H]$. 

We find a MPS approximation to the purification~\eqref{eq:puri} in the thermodynamic limit using the iTEBD algorithm to approximate $\exp[-(\beta/2) H]$.
The transfer operator of this MPS determines the norm of the state (and can be chosen to be normalized by appropriate gauge choice). In order to determine the density $s_2$, we thus just need the dominant eigenvalue $\eta$ of the transfer operator of  $\rho_{\beta}^2$, $\mathbb E_{\rho_\beta^2}$,\footnote{See Appendix~\ref{app:transferop} for details on the transfer operator.}
\begin{equation} \label{eq:s2_eta}
    s_2 = -\ln\eta .
\end{equation}
The value of $\eta$ can be determined efficiently by considering the action of $\mathbb E_{\rho_\beta^2}$ on an eigenvector and using iterative eigensolvers. 

For our meson studies we choose exemplarily the Ising QFT parameter point specified by $M_h/M_g \approx 0.09$ in the nonintegrable ferromagnetic phase. Additionally, we vary the individual masses $M_h$ and $M_g$ according to the parametrization
\begin{equation} \label{eq:Mhg_params}
    M_{h,g}^{(n)} = \frac{M_{h,g}^{(0)}}{2^n} ,
\end{equation}
where $M_h^{(0)}/J=0.125$, $M_g^{(0)}/J\approx 1.356$ and the ratio $M_h/M_g$ is kept constant. For increasing values of $n$, the masses are decreasing, i.e.\ we approach the critical point in the phase diagram. This means nothing else than taking the scaling limit previously described in section~\ref{sec:prelim_Ising}. The resulting masses $M_1/J$ of the first meson, which we read out from \cite{Fonseca:2006au}, are hence also decreasing. For a fixed range of lattice spacing values, $\beta J \in [1,16]$, we therefore can probe different physical temperatures, measured in units $\beta M_1$.

Before discussing the results of these simulations, let us consider the expected scaling behavior in the transverse field Ising model, i.e.\ without longitudinal field perturbations causing confinement. 
In this case the model can be mapped to a free fermion system (see e.g.~\cite{sachdev_2011}) from which the partition function follows as
\begin{equation}
    Z_\beta = \Tr\left[ \e^{-\beta H} \right] = \prod_k 2 \cosh\left( \frac{\beta \epsilon_k}{2} \right) ,
\end{equation}
where the single-particle energies are given by $\epsilon_k = 2J \sqrt{1 + h^2 - 2h\cos(k)}$ for discrete wave number $k$. The definition \eqref{eq:s2} implies immediately that the second R\'enyi entropy density can be calculated as $s_2 = 2\beta[f(2\beta)-f(\beta)]$ from the free energy density $f(\beta) = -\frac{1}{N \beta} \ln Z_\beta$. In the thermodynamic limit, this leads to
\begin{equation}
    s_2(\beta) = \ln 2 - \frac{1}{\pi} \int_0^\pi \mathrm dk \ln\left[ \frac{\cosh(\beta\epsilon_k)}{\cosh^2(\frac{\beta\epsilon_k}{2})} \right] .
\end{equation}
At low temperatures, i.e.\ in the limit $\beta\to\infty$, and away from criticality the dominant contribution to $s_2$ originates from $k=0$, which gives
\begin{equation} \label{eq:s2_lowT}
    s_2 \sim \frac{\e^{-2\beta J\vert h-1\vert}}{\sqrt{\frac{\pi\beta J h}{\vert h-1\vert}}} .
\end{equation}
Note that the term $\vert h-1\vert$ is proportional to the free fermion mass $M_h = 2J\vert 1-h\vert$. On the other hand, at the critical point ($h=1$), the integral can be evaluated to give a power-law decay with the inverse temperature,
\begin{equation} \label{eq:s2_highT}
    s_2 \sim \frac{\pi}{16} \left( \frac{1}{\beta J}+\frac{1}{16(\beta J)^3}+\ldots \right) .
\end{equation}
The $1/\beta$ behavior is the (free) CFT result and the $1/\beta^3$ correction is a lattice effect.

The exponential scaling at low temperatures in the massive free fermion regime is clearly distinct from the power-law decay at high temperatures at criticality. While we do not have a prediction for the scaling in the nonintegrable regime, it is tempting to assume that the temperature dependence could be equally determined by the existence of a mass gap in the spectrum. To test this hypothesis, we need to analyze both the low-temperature regime, given for $\beta M_1 \gg 1$, and the high-temperature regime, given by $\beta M_1 \lesssim 1$. In the latter case, we additionally need to ensure that the result is not dominated by lattice excitations, i.e.\ $\beta J \gtrsim 1$.

Fig.\,\ref{fig:s2_lowT_scaling} (left panel) shows the behavior of $s_2$ as a function of the inverse temperature $\beta M_1$. In all cases, we observe an exponential decay at low temperatures. In the right panel, we analyze this regime quantitatively (exemplified for $n=1$) by applying a signal analysis technique based on the Prony method onto $s_2$ as a function of $\beta J$ and decomposing it into harmonic contributions with complex parameters.\,\footnote{We refer to Appendix~\ref{app:prony} for more details on the method.} Two clear poles on the imaginary axis are visible. They quantify the exponential decay at low temperatures. The values agree with the masses of the first two meson states (shown as black circles). Based on these findings, we expect the following low-temperature scaling of the R\'enyi entropy density,
\begin{equation} \label{eq:exponential_damping}
    s_2 \sim \sum_i c_i \e^{-\beta M_i} ,
\end{equation}
for some coefficients $c_i$ and meson masses $M_i$, whereby the lowest mass $M_1$ is obviously dominating. This form of the asymptotic scaling is the natural generalization of the massive free fermion result $s_2 \sim \e^{-\beta M_h}$ in \eqref{eq:s2_lowT}.

Fig.\,\ref{fig:s2_highT_scaling} (top panel) shows again $s_2$ as a function of $\beta M_1$, but emphasizing the high-temperature regime in a double-logarithmic plot. To the left of the gray line the high-temperature regime $\beta M_1 \lesssim 1$ starts, while the dashed lines denote the estimated lattice scale for each curve. As $n$ is increasing, i.e.\ the individual masses $M_{h,g}$ are decreasing, one can observe the emergence of a linear scaling in between these two curves, indicating a power-law behavior. Assuming the general functional dependence 
\begin{equation} \label{eq:power_law_scaling}
    f(\beta) \sim \beta^{-p} ,
\end{equation}
we extract in Tab.\,\ref{tab:Renyi_scaling} the values of the power-law exponents $p$ from a linear fit to the numerical data in between these two scales. For $f = s_2$, the value $p=1$ is obtained with a very good accuracy of only a few percent. This implies that the high-temperature scaling behavior matches the CFT expectation from \eqref{eq:s2_highT}. Based on the clear observation from Fig.~\ref{fig:s2_lowT_scaling} of meson states in the entropy scaling at low temperatures and their absence at high temperatures, we interpret this signature as strong evidence for meson melting. In this regime, which is not yet influenced by the lattice scale, melted meson states do not leave an imprint on the entropy scaling any more, which is why the scaling is identical to a critical system. To further corroborate this finding, we additionally analyze also the scaling behavior of the thermal energy density (bottom panel in Fig.\,\ref{fig:s2_highT_scaling}) and its derivative. Both these thermodynamic quantities show a similar power-law scaling at high temperatures. The corresponding exponents are tabulated in Tab.\,\ref{tab:Renyi_scaling}. At $n=3$ they agree well with the analytical expectation, which is set by the scaling dimension $\Delta=2$ of the associated CFT operator. Interestingly, the R\'enyi entropy $s_2$ seems to allow an even better estimation of the scaling exponent for all values of $n$, indicating the potential of entropic quantities to identify the meson melting process.

In summary, we have shown that the scaling behavior of the second R\'enyi entropy density provides a clear signature of the asymptotic high- and low-temperature regime in a nonintegrable Ising QFT with meson bound states. At low temperatures, we found that the behavior is dominated by the exponential damping of the form \eqref{eq:exponential_damping}. At high temperatures, the power-law dependence \eqref{eq:power_law_scaling} matches the CFT behavior. We interpret the latter scaling as signaling the fact that the meson states have been melted. In fact, a similar analysis for the E$_8$ regime yields the same qualitative picture, providing evidence that the results are independent from integrability properties of the underlying system. Naively, it might look trivial that a massive QFT matches at high temperatures the CFT behavior due to the fact that the temperature then provides effectively the only scale in the system. (A similar situation for the massive free fermion regime in the Ising model is discussed for example in \cite{sachdev_2011}.) However, in our present studies we do have nonperturbative meson bound states in the spectrum, which are ``visible'' at low temperatures, and we observe a smooth transition to the high-temperature regime, which we can disentangle from the UV lattice scale. We therefore see the scaling properties of the second R\'enyi entropy as a clear witness of the meson melting process in a static situation. While we have focused on the second order R\'enyi entropy, it is possible that this conclusion holds also for higher-order ones.

\section{Results from dynamics after thermal quenches}
\label{sec:quench}

\subsection{Setup}

We consider quenches from the thermal equilibrium state of a pre-quench Hamiltonian $H_0$. At time $t=0$ an instantaneous global quench is applied and the system evolves with Hamiltonian $H_1$. In particular, we construct the purification $\sqrt{\rho(0)} \equiv \e^{-\beta H_0/2}\ket{\Phi}$ in the form \eqref{eq:puri} and simulate its time evolution $\sqrt{\rho(t)} \equiv U(t)\sqrt{\rho(0)}$, where $U(t) = \e^{-itH_1}$ is the quenched time evolution operator. The thermal state approximation is then given by $\rho(t) \equiv U(t)\sqrt{\rho(0)}\sqrt{\rho(0)}^\dagger U^\dagger(t)$.\,\footnote{Note that this choice of purification is not unique: any unitary applied to the ancillary sites results in the same reduced state. In particular, it may be possible to find at any time unitaries $V$ such that the purification $U(t)\sqrt{\rho(0)}V^{\dagger}$ has lower entropy~\cite{Hauschild2018}. We fix our purification as $U(t) \sqrt{\rho(0)}$ because we are interested in singling out the dynamical properties of the Hamiltonian $H_1$.} An MPO representation of this principle is illustrated in panel (a) of Fig.\,\ref{fig:QuenchTypes}.\,\footnote{Details of the diagrammatic TN notation are presented in Appendix~\ref{app:transferop}.} We fix for $H_1$ the parameters that correspond to $M_h^{(0)}$ and $M_g^{(0)}$ from the parametrization \eqref{eq:Mhg_params} (namely $h=0.9375$ and $g\approx 0.0746$), as this corresponds to the smallest time period $2\pi J/M_1 \approx 4.2$ (for $M_1/J \approx 1.5$) of the first meson, and we know, from our earlier equilibrium studies \cite{Banuls:2019qrq}, that in this regime, MPO simulations and Prony signal analysis~\cite{peter2014generalized} allowed a reliable identification of the QFT meson masses.

The previously considered global R\'enyi entropy density $s_2$ in Eq.~\eqref{eq:s2_eta} is constant under unitary  time evolution, and computing its corresponding value for a finite subsystem is computationally too demanding for large bond dimensions. 
However, we instead can study \textit{reflected entropies}, which we introduced previously in section~\ref{subsec:entropies}, in the following way. Since we have a MPO approximation of the purification as well as its time evolution available, the entropies are directly accessible from the Schmidt values of the ansatz.
In particular, if $\{\lambda_\alpha \}$ are the Schmidt values across a cut of the purified state $\sqrt{\rho(t)}$, the von Neumann and 2-R\'enyi reflected entropies are given by
\begin{align}
    \tilde s_1 &= -\sum_{\alpha=1}^{\chi} \lambda_\alpha^2 \ln(\lambda_\alpha^2) , \label{eq:s_1}\\ 
    \tilde s_2 &= -\ln\left( \sum_{\alpha=1}^{\chi} \lambda_\alpha^4 \right) .\label{eq:s_2}
\end{align}
The thermal approximation is known to be efficient, and the error of the time-evolved MPO can be monitored by the truncation. We therefore can accurately estimate the targeted entropies. Using this procedure, we study in this section the growth and oscillatory behavior of these quantities after a thermal quench.

\begin{table}[t]
\centering
\caption{\label{tab:effective_temps} Comparison of the pre-quench temperatures $\beta J$ of the initial thermal state (w.r.t.\ $H_0$) and the resulting post-quench effective temperatures $\beta^* J$ and $\beta^* M_1$ (w.r.t.\ $H_1$) for the different quench types shown in Fig.\,\ref{fig:QuenchTypes}.}
\begin{tabular}{ c | c | c | c | c | c | c } 
     \hline
initial temperature $\beta J$ & 
\multicolumn{6}{c}{effective temperature $\beta^* J$, ($\beta^*M_1$)}\\ 
      & \circled{1}  & \circled{2}  & \circled{3}  & \circled{4}  & \circled{5}  & \circled{6}  \\ \hline 
 16   & 6.10  & 3.55  & 1.55  & 0.91  & 3.55  & 0.93  \\
      & (9.3) & (5.4) & (2.4) & (1.4) & (5.4) & (1.4) \\
 2    & 2.00  & 1.92  & 1.34  & -     & -    & -      \\
      & (3.0) & (2.9) & (2.0) &       &      &        \\
 0.97 & -     & -     & 0.91  & -     & -    & -      \\
      &       &       & (1.4) &       &      &        \\
 0.5  & 0.50  & -     & -     & -     & -    & -      \\
      & (0.8) &       &       &       &      &        \\ \hline
\end{tabular}
\end{table}

\begin{figure}[t]
    \centering
   \begin{minipage}{\columnwidth}
   {\raggedright (a)\\} 
   \includegraphics[width=0.75\columnwidth]{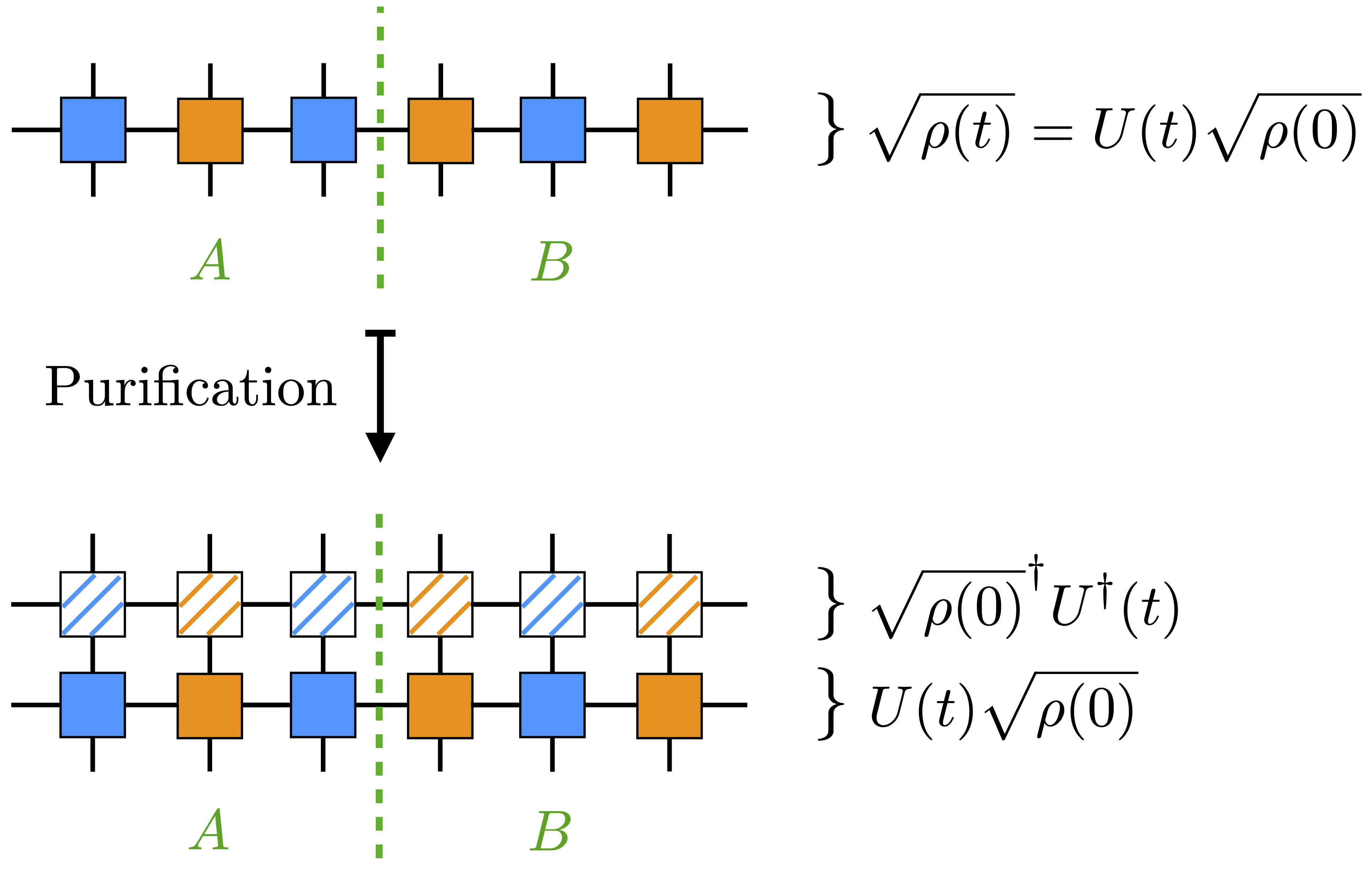}
   \end{minipage}
   \begin{minipage}{\columnwidth}
   {\raggedright (b)\\}
   \includegraphics[width=0.65\columnwidth]{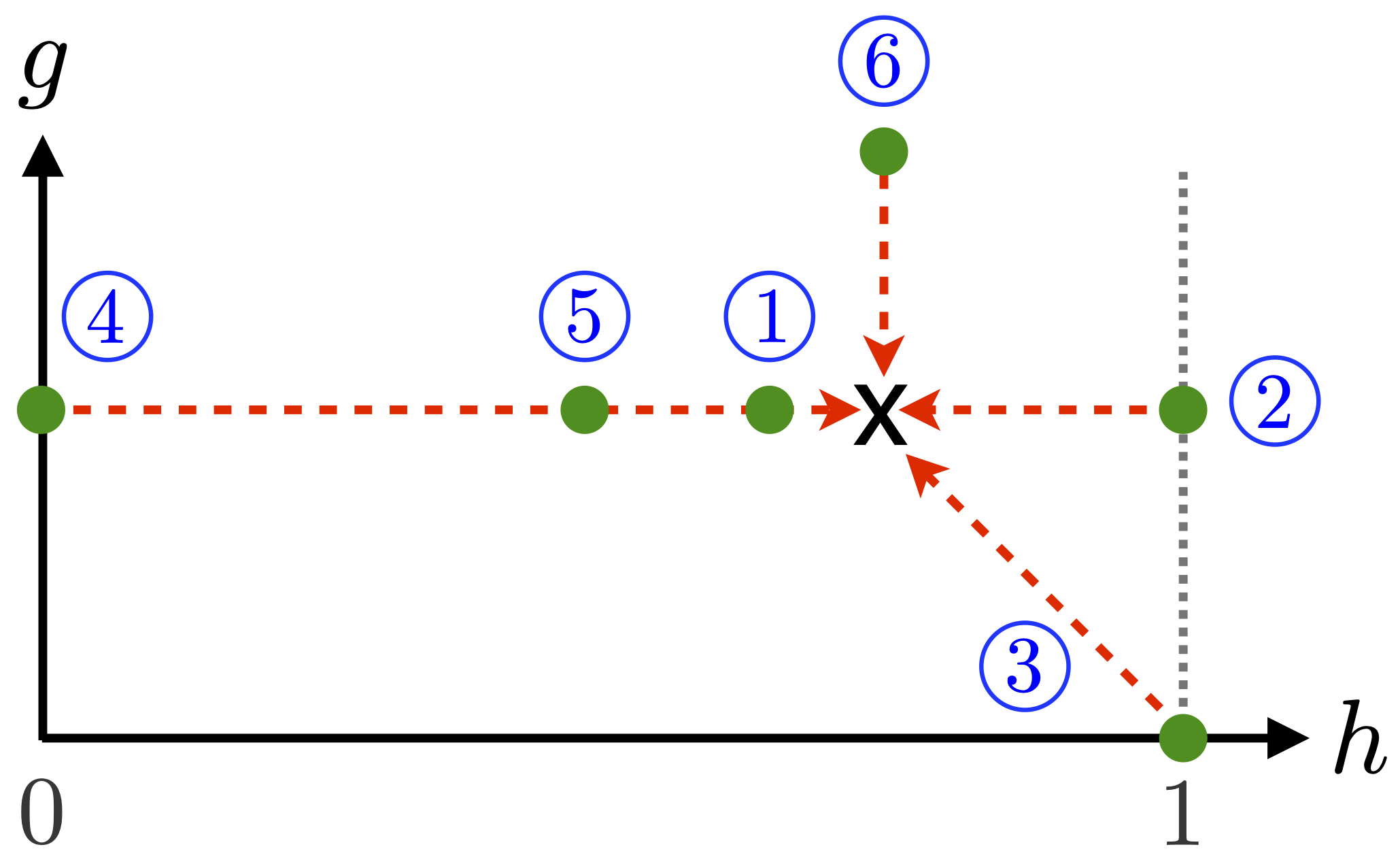} \\ 
   \end{minipage}
    \caption{Overview of the TN setup (a) and different quench protocols in the transverse ($h$) vs.\ longitudinal ($g$) field plane (b). The E$_8$ regime is marked by the gray dotted line. Thermal initial states $\sqrt{\rho(0)}$ are prepared at inverse temperature $\beta$ for the parameter locations marked by green dots. By quenching to the final nonintegrable ferromagnetic point, denoted by red arrows towards the black cross, an effective temperature $\beta^*$ as listed in Tab.~\ref{tab:effective_temps} is induced by the thermal quench. Entanglement measures are then calculated for a semi-infinite bipartition into a subsystem $A$ and its complement $B$ for the purification of the time-evolved state $U(t)\sqrt{\rho(0)}\sqrt{\rho(0)}^\dagger U^\dagger(t)$.}
    \label{fig:QuenchTypes}
\end{figure}

\begin{figure*}[t]
    \centering
   \includegraphics[width=0.49\textwidth]{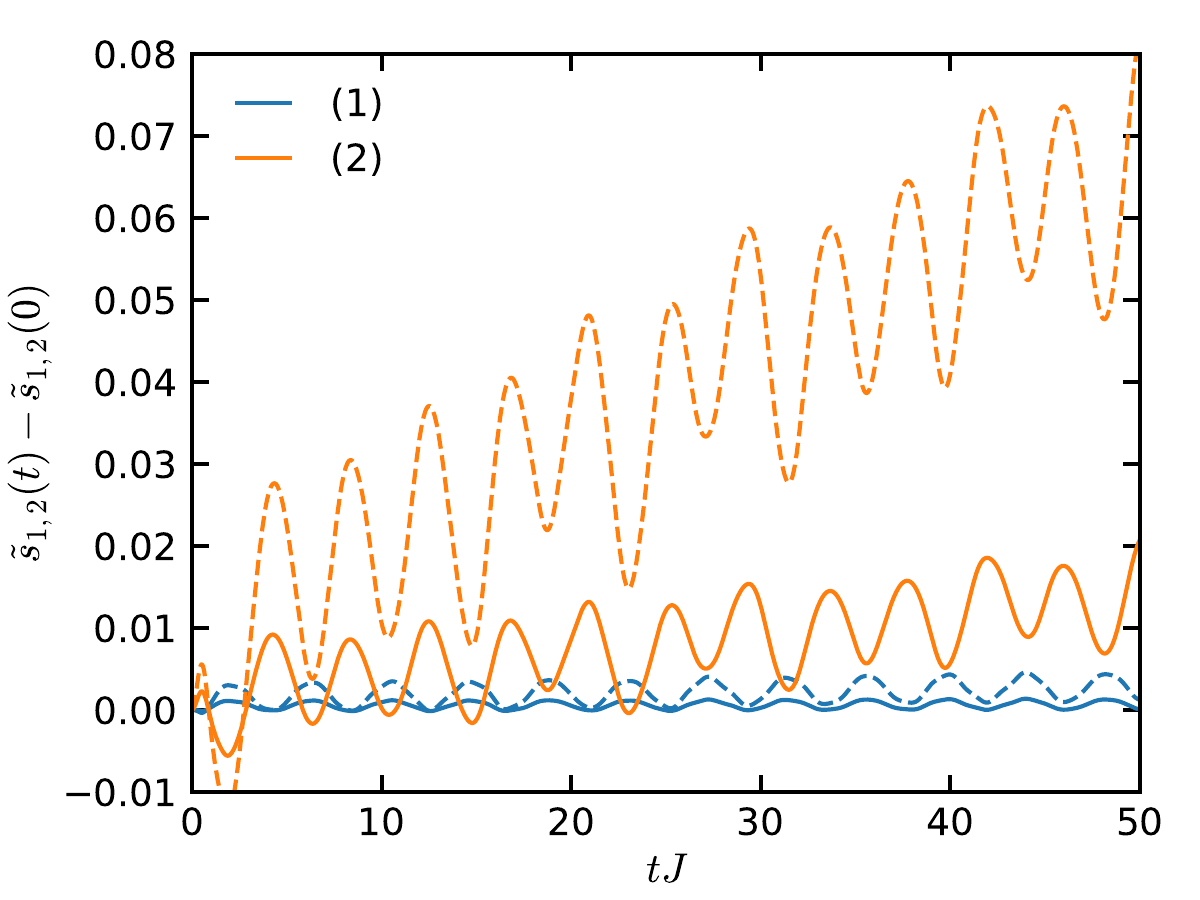} 
   \includegraphics[width=0.49\textwidth]{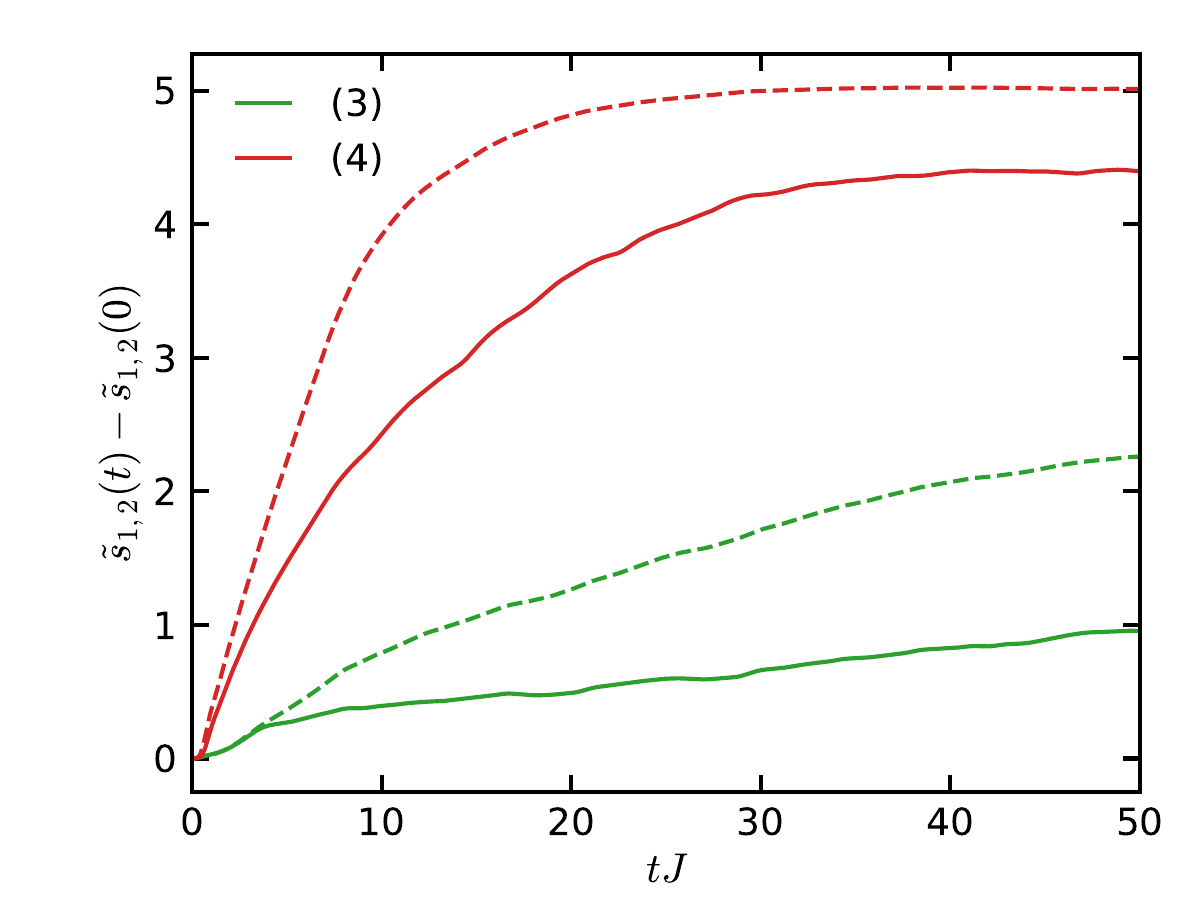} 
    \caption{Time dependence of the first (dashed curves) and second (solid curves) reflected entropies for the quench protocols (1)-(4) (as defined in Fig.\,\ref{fig:QuenchTypes}) with an initial temperature $\beta J=16$. With increasing effective temperature (in order of the quench labels) the entanglement growth is getting enhanced and suppresses the entanglement oscillations. The plateau arising in quench type (4) at late times is an unphysical effect caused by the truncated bond dimension (see main text for further discussions).}
    \label{fig:s12_overview}
\end{figure*}

After the quench, the state is not in thermal equilibrium for $H_1$. However, we can define an \textit{effective temperature}, which corresponds to the thermal equilibrium of $H_1$  at the same (conserved) energy density of the system. The effective inverse temperature $\beta^*$ is then given in terms of the pre- and post-quench Hamiltonian densities $\bar H_0$ and $\bar H_1$ by the condition
\begin{equation} \label{eq:betaeff}
    \frac{\Tr\left[ \bar H_1 \e^{-\beta \bar H_0} \right]}{\Tr \left[\e^{-\beta \bar H_0}\right]} = \frac{\Tr\left[ \bar H_1 \e^{-\beta^* \bar H_1} \right]}{\Tr\left[ \e^{-\beta^* \bar H_1}\right]} .
\end{equation}
Since we keep the final Hamiltonian $H_1$ fixed, the effective temperature can be increased either by varying the initial temperature $\beta$ w.r.t.\ a fixed pre-quench Hamiltonian $H_0$ or by modifying the parameters of $H_0$ for fixed $\beta$. We will employ both methods in our studies. In the latter scenario, we choose the initial points in the phase diagram as shown in Fig.\,\ref{fig:QuenchTypes} [panel(b)]. The different quench protocols, labeled by the numbers in the plot, result in the effective temperatures $\beta^*J$ (measured w.r.t.\ $H_1$) summarized in Tab.\,\ref{tab:effective_temps} (determined up to two digits) for several initial temperatures $\beta J$ (measured w.r.t.\ $H_0$). The quench Hamiltonian $H_1$ corresponds to the fixed parameters $h=0.9375$ and $g \approx 0.0746$. In the quench protocol \circled{1}, we have $h = 0.93$, and for \circled{5} $h = 0.8732$. In type~\circled{6}, the longitudinal field is chosen asymptotically large as $g = 100$. The remaining initial parameters are identifiable from Fig.\,\ref{fig:QuenchTypes}. Note in particular that protocol \circled{4} starts in the classical regime, whereas \circled{2} is in the E$_8$ phase and \circled{3} at the critical point. The different quench types encompass pure transverse quenches (\circled{1},\circled{2},\circled{4},\circled{5}) and longitudinal quenches \circled{6}. Otherwise, both fields are quenched. We consider the case of an initial low temperature (i.e.\ close to the ground state) at $\beta J=16$ as well as higher initial temperatures, cf.\ Tab.\,\ref{tab:effective_temps}. Note that protocol \circled{5} is tuned such that is has the same effective temperature as \circled{2} for $\beta J=16$. Similarly, types \circled{4} and \circled{6} at $\beta J=16$ and \circled{3} at $\beta J=0.97$ result in (nearly) the same effective temperature. 
Overall, we have adjusted our setup to achieve both low physical temperatures $\beta^* M_1 \gg 1$ as well as high temperatures $\beta^* M_1 \lesssim 1$, while avoiding the lattice regime $\beta^* J \ll 1$.\,\footnote{Note that the effective temperatures $\beta^*J=2.00$ and $\beta^*J=0.50$ for type (1) mean that the difference to the initial $\beta J$ is smaller than our considered precision of two digits.}

In the following, we ensure that our simulations represent faithful physical results by checking the convergence of $\tilde s_1$ and $\tilde s_2$ with the maximally allowed value $\chi$ of the bond dimension in \eqref{eq:s_1} and \eqref{eq:s_2}. This becomes relevant if there is a large entanglement growth, which potentially could not be captured by the TN ansatz. Moreover, we again also explicitly consider the scaling limit in which the 1+1$D$ Ising QFT emerges from the discrete Ising model. We therefore can draw reliable interpretations from our described quench setup.

\subsection{Effective temperature effects}

We start with the analysis of quenches from an initial state close to the ground state of $H_0$, prepared at low temperature $\beta J=16$. Fig.\,\ref{fig:s12_overview} shows the results for $\tilde s_1$ (dashed curves) and $\tilde s_2$ (solid curves) as a function of time (in units of $J$) for the quench protocols \circled{1}-\circled{4}. As listed in Tab.\,\ref{tab:effective_temps}, the effective temperatures are in increasing order from $\beta^*M_1 \approx 9.3$ to $\beta^*M_1 \approx 1.4$. As can be appreciated from Fig.\,\ref{fig:s12_overview}, the reflected entropy grows faster for larger effective temperature. For our discussion, we limit ourselves to regimes which are numerically achievable with our MPO simulations. When the entanglement grows fast, this is possible only up to a certain time, after which the observed entropy saturates due to a finite bond dimension effect (see e.g. Fig.~\ref{fig:s12_overview} after $tJ\approx 20$). Thus our analysis is limited to early and intermediate time scales.

In the quench types \circled{1} and \circled{2}, which are shown in detail in the left column, $\tilde s_1$ and $\tilde s_2$ exhibit an oscillatory behavior, similar to the ground state case first discussed in \cite{kormos2017real}. As we will analyze in detail below, these entanglement oscillations are caused by the QFT meson states at characteristic frequencies corresponding to their masses. Type (2) (orange curves) shows on top of these entanglement oscillations an overall increase. This can be explained in the quasiparticle model either through the production of mesons with finite velocities, or the fact that the arising envelope frequencies (that bound the entanglement growth when mesons are produced at rest) would require (much) longer time scales. The latter case is plausible when keeping in mind that we have chosen a parameter point in the QFT regime of the Ising model close to its critical point. The meson masses, e.g.\ in absolute units of $J$, are smaller as compared to a semiclassical regime and hence result in much longer time recurrences. From Fig.\,\ref{fig:s12_overview} it becomes apparent that the entanglement oscillations are heavily suppressed as the effective temperature is raised from type \circled{1} to \circled{4}. This indicates that the meson states do not dominate the entanglement growth at their characteristic mass frequencies anymore. In particular, the first reflected entropy of quench protocol \circled{4} (red dashed curve in the right panel) grows fast at early times, with an apparently linear behavior. Below, we will study each of these effects in detail.

\begin{figure*}[t]
    \centering
\begin{minipage}{0.32\textwidth}
   \includegraphics[width=\textwidth]{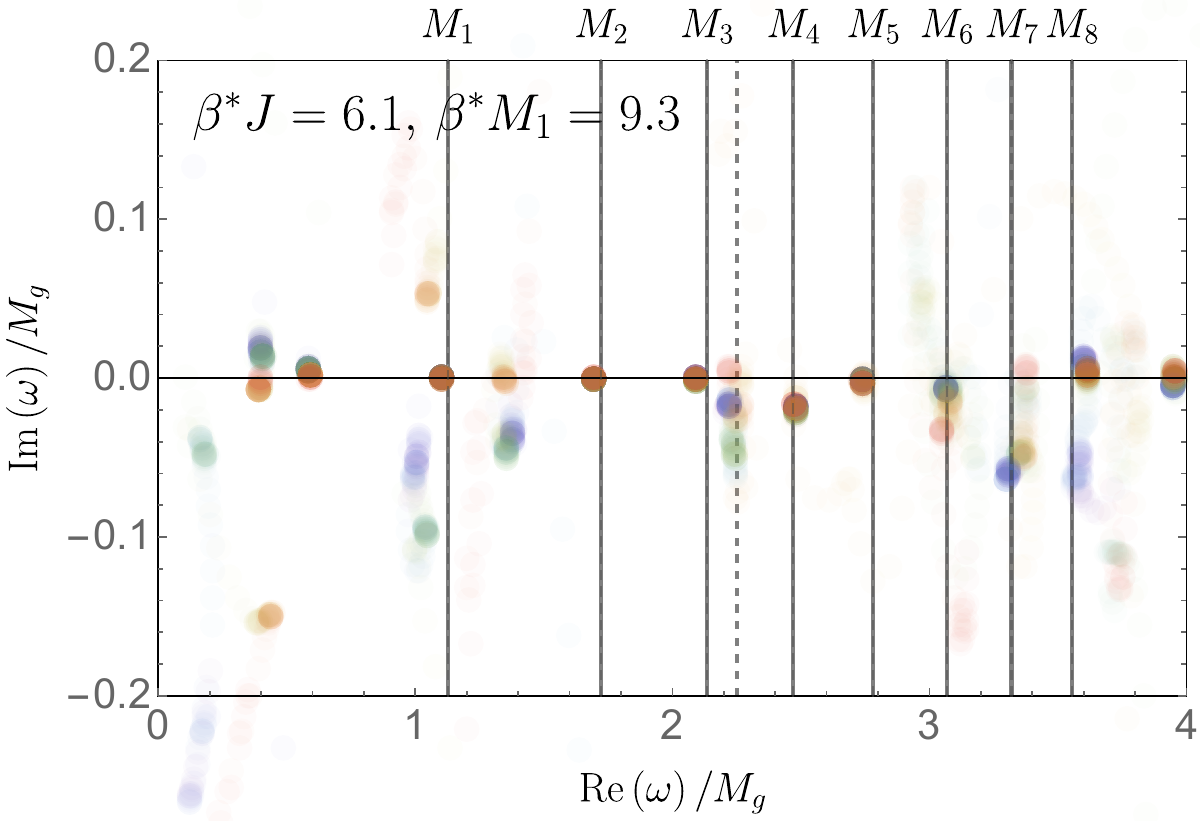} 
\end{minipage}
\begin{minipage}{0.32\textwidth}
   \includegraphics[width=\textwidth]{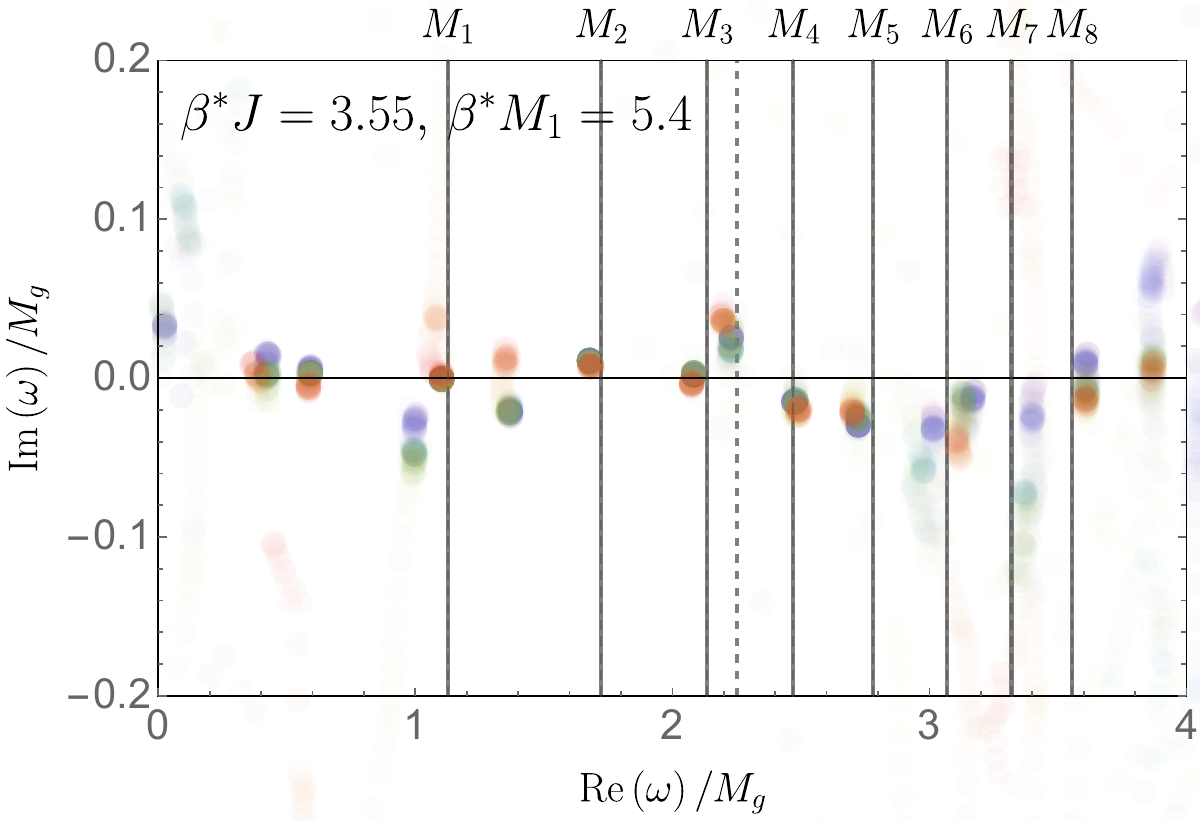} 
\end{minipage}
\begin{minipage}{0.32\textwidth}
   \includegraphics[width=\textwidth]{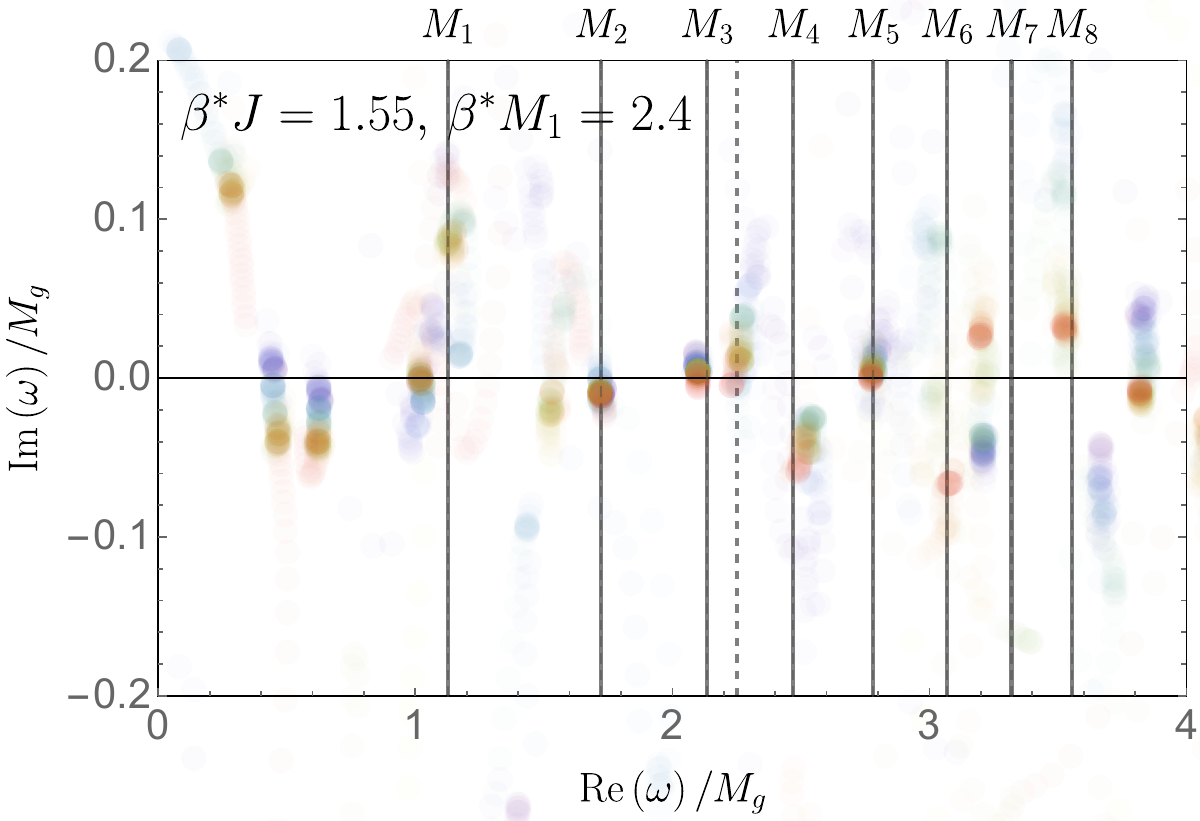} 
\end{minipage}
\begin{minipage}{0.32\textwidth}
   \includegraphics[width=\textwidth]{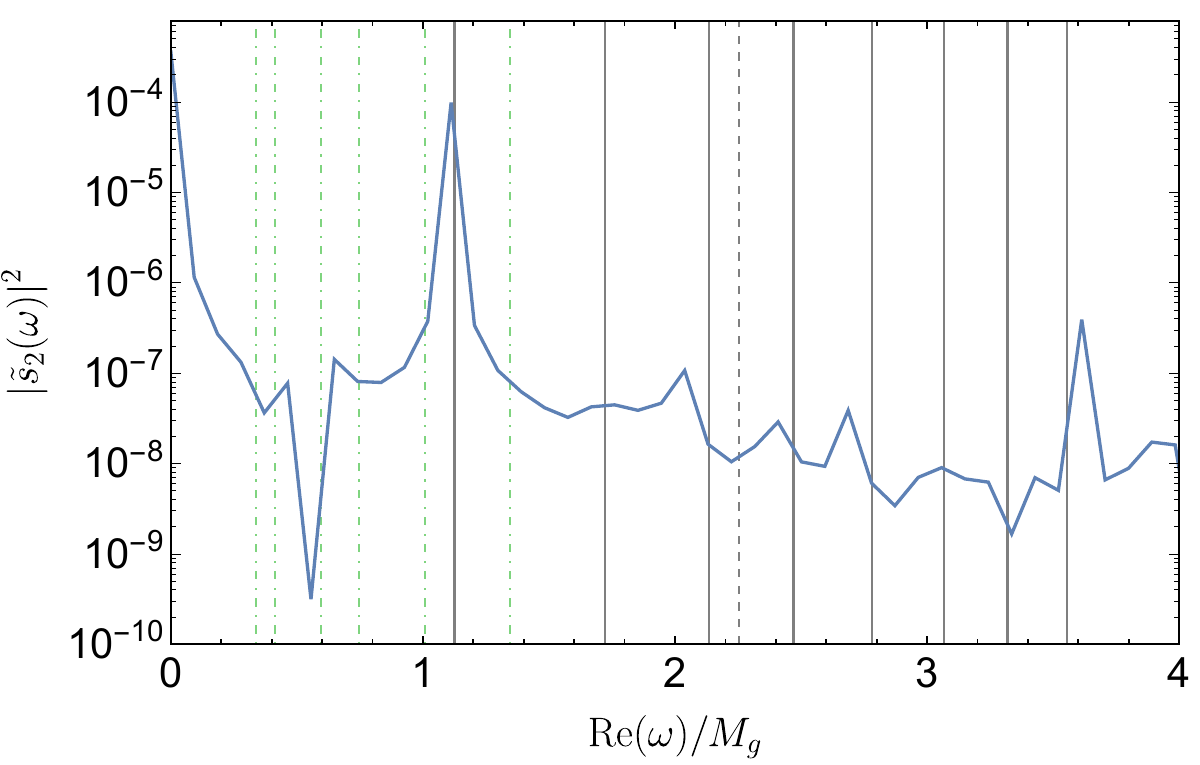}\\ 
   \circled{1}
\end{minipage}
\begin{minipage}{0.32\textwidth}
   \includegraphics[width=\textwidth]{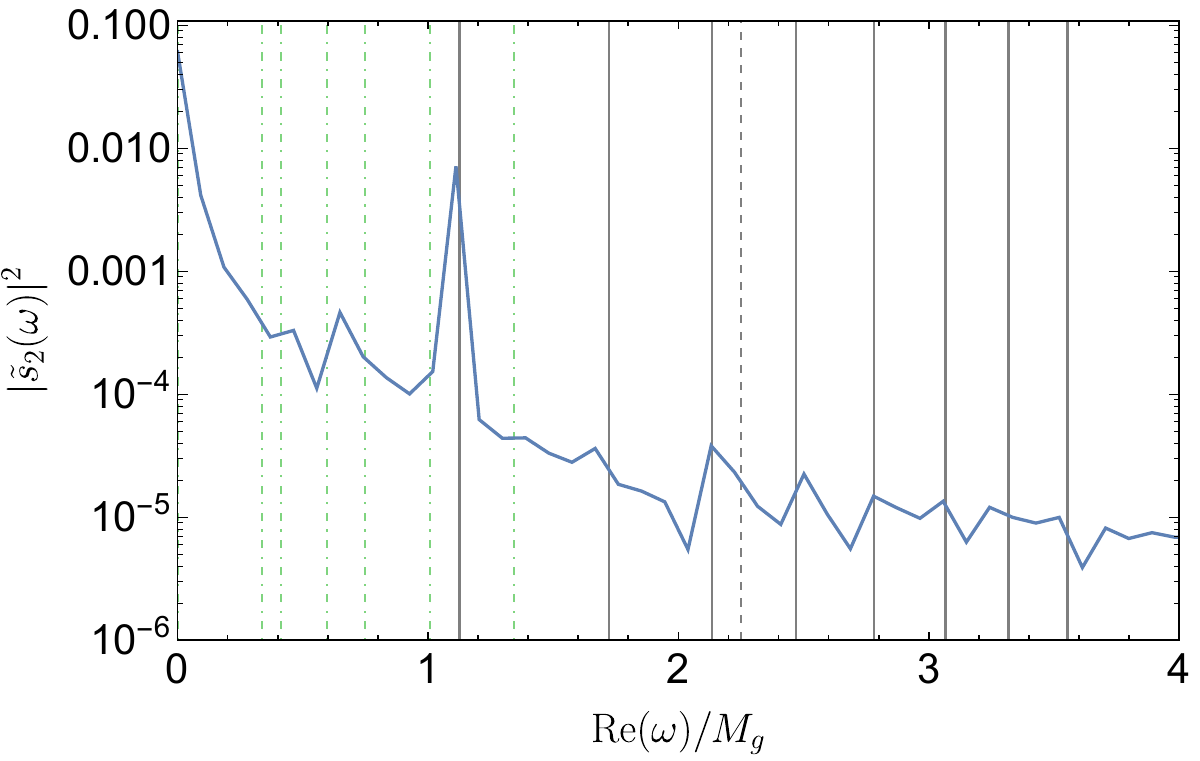}\\ 
   \circled{2}
\end{minipage}
\begin{minipage}{0.32\textwidth}
   \includegraphics[width=\textwidth]{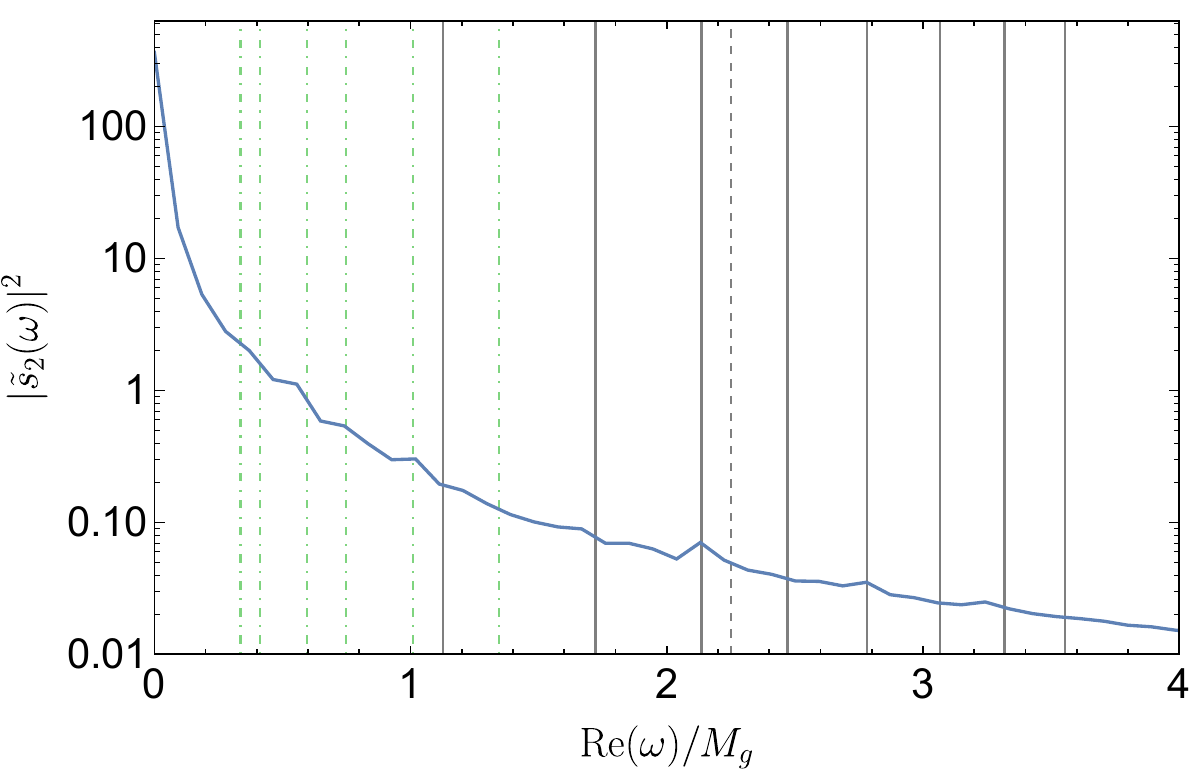}\\ 
   \circled{3}
\end{minipage}
    \caption{Results of the Prony signal analysis (top) and Fourier spectra (bottom) of $\tilde s_2$ for the quench protocols (1)-(3) (from left to right) with an initial temperature $\beta J=16$ and effective temperatures $\beta^*M_1$ as shown. Solid vertical lines represent meson masses from ref.~\cite{Fonseca:2006au} and dashed lines mark the continuum threshold of $2 M_1$. Green dash-dotted lines mark all possible mass differences of the first 4 mesons. Both methods allow an identification of meson states as poles in the complex frequency plane or peaks in the Fourier spectrum. When the effective temperature is raised, meson poles become fuzzier, while the spectrum increases and flattens out.
    }
    \label{fig:s2_FT_and_Prony_beta16}
\end{figure*}

These findings are in general valid for both $\tilde s_1$ and $\tilde s_2$ as visible in Fig.~\ref{fig:s12_overview}. Observe, however, that the growth of $\tilde s_2$ is suppressed as compared to $\tilde s_1$. For this reason, it turns out that the second reflected entropy allows for a more precise identification of meson masses from the entanglement oscillations when evaluating both the Prony signal analysis method as well as the Fourier spectra. That is why we consider $\tilde s_2$ for a detailed analysis in the following discussion. The results are presented in Fig.\,\ref{fig:s2_FT_and_Prony_beta16} for protocols \circled{1}-\circled{3} (from left to right). At the lowest effective temperature $\beta^*M_1 \approx 9.3$ for quench type \circled{1} (left column), the Prony analysis allows a clear identification of the first 5 meson poles, which are in good agreement with their QFT mass values when compared to the predictions in \cite{Fonseca:2006au} (solid vertical lines). Additional features are visible also at $M_6$ to $M_8$, which, however, become fuzzier and hence more uncertain. In addition, the continuum threshold at $2M_1$ is identifiable as a vertical line of poles, indicating a branch cut (shown as the dashed vertical line). In the corresponding Fourier spectra, the meson poles translate into peaks at their frequency values. The peak of the first meson is largely dominating over the other ones. Overall, the Fourier spectrum is decreasing over several orders of magnitude towards larger frequencies. The green dash-dotted lines in the Fourier spectra mark all 6 possible mass differences between the first 4 meson states. At their respective values, kinks appear in the spectrum, corresponding to additional poles in the Prony plot. They lie at frequencies smaller than the first meson mass as well as between $M_1$ and $M_2$. 

\begin{figure}[t]
    \centering
  \includegraphics[width=0.49\textwidth]{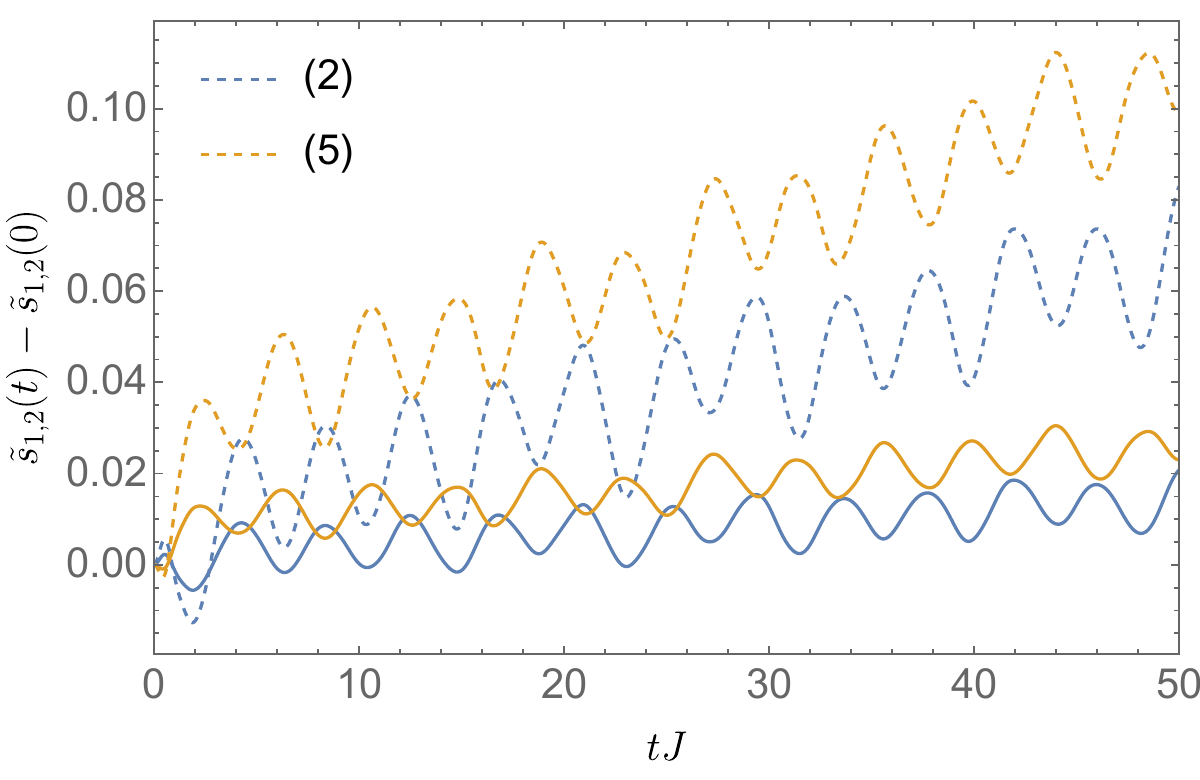} 
  \includegraphics[width=0.49\textwidth]{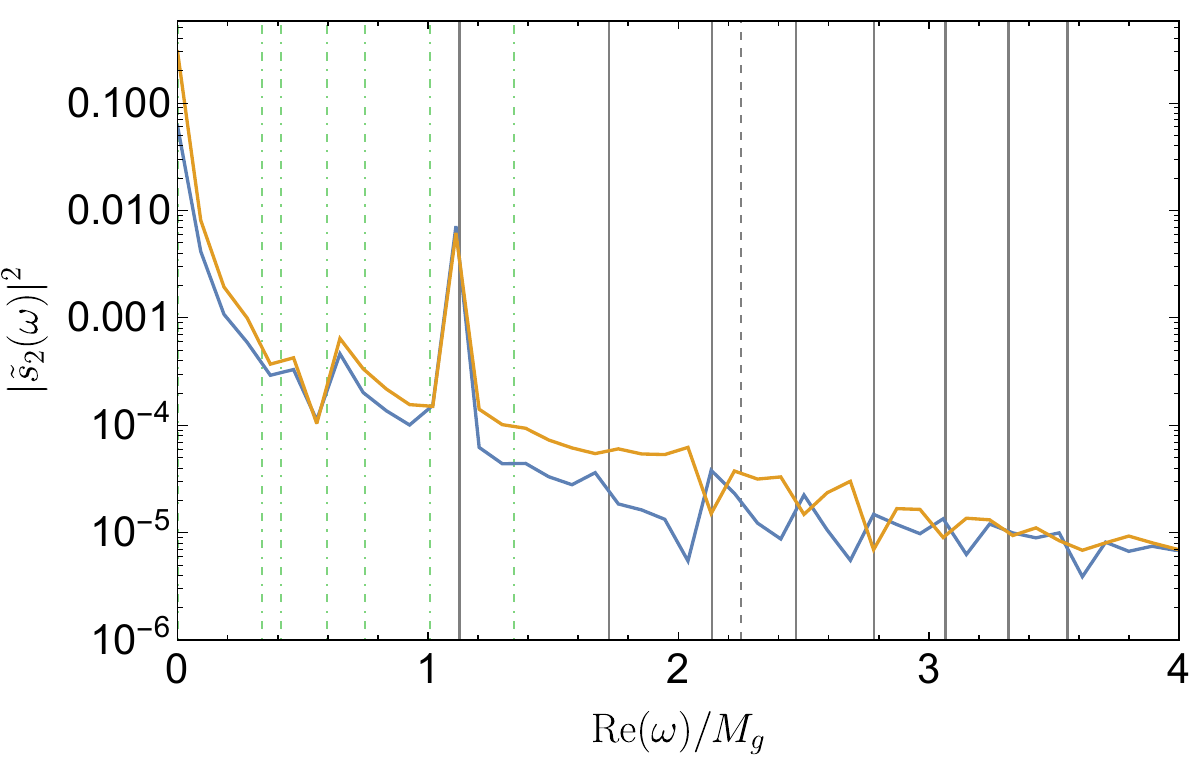} 
    \caption{Comparison of the quench protocols (2) (blue curves) and (5) (yellow curves) for an initial temperature $\beta J=16$ and the same effective temperature $\beta^* M_1 \approx 5.4$.
    Top: Time dependence of the first (dashed) and second (solid) reflected entropy in both models. Bottom: Corresponding Fourier spectra of the second reflected entropy. The different regime of the initial state (E$_8$ vs.\ nonintegrable ferromagnetic) causes only mild differences in the behavior of entanglement growth and oscillations.}
    \label{fig:comparison_type25}
\end{figure}

As the effective temperature is raised to $\beta^*M_1 \approx 5.4$ in protocol \circled{2} (middle column) and $\beta^*M_1 \approx 2.4$ in protocol \circled{3} (right column), the meson pole identifications become fuzzier in the Prony analysis. Notably, the absolute values of the corresponding Fourier spectra are raised by 3 and 6 orders of magnitude, respectively. For protocol \circled{3}, the Fourier spectrum is flattening out (relative to the overall magnitude). Since in all cases the post-quench nonintegrable QFT regime is identical, we attribute this effect to the increasing effective temperature. This hypothesis is corroborated by data from a different quench with the same effective temperature (see Fig.~\ref{fig:comparison_type25}). When comparing the meson peak heights to the overall magnitude of the Fourier spectrum, the impact of the meson mass frequencies seems to get suppressed at high temperatures. We see this thermal suppression effect on the Fourier spectrum as a first hint of how the melting of meson states at even higher temperatures can be detected. A sequential process in similarity to QCD (cf.\ the discussion in section~\ref{sec:prelim_pheno}) could show up in this description through a stepwise decrease and disappearance of individual meson peaks in the Fourier spectrum of the reflected entropies. The available data in Fig.\,\ref{fig:s2_FT_and_Prony_beta16} are, however, inconclusive about whether this process is consistent with this picture.

To analyze the influence of the initial state in more detail, we compare quenches from different initial states at the same effective temperature. Fig.~\ref{fig:comparison_type25} shows the time evolution of the reflected entropies and resulting Fourier spectra of quench protocols \circled{2} and \circled{5} starting from a pre-quench thermal state at $\beta J=16$. Both states have the same effective temperature $\beta^*J \approx 3.55$ or $\beta^*M_1 \approx 5.4$, but are initially prepared in the E$_8$ regime versus the nonintegrable ferromagnetic phase. The figure shows that $\tilde s_1$ (dashed curves) and $\tilde s_2$ (solid curves) differ only marginally in the two quench protocols and exhibit a slight phase shift. The first meson peak dominates the Fourier spectrum equally in both models. For the available frequency resolution and range, both Fourier spectra are in close correspondence with each other. These results indicate that the described features, at least for this particular case, are primarily driven by thermal effects, set by the effective temperature, while the influence of the initial state seems subdominant. Below, we will find a similar situation also for other models at even higher temperatures, providing evidence for the robustness of the physical interpretation about thermal effects from our findings.

\begin{figure*}[t]
    \centering
\begin{minipage}{0.32\textwidth}
   \includegraphics[width=\textwidth]{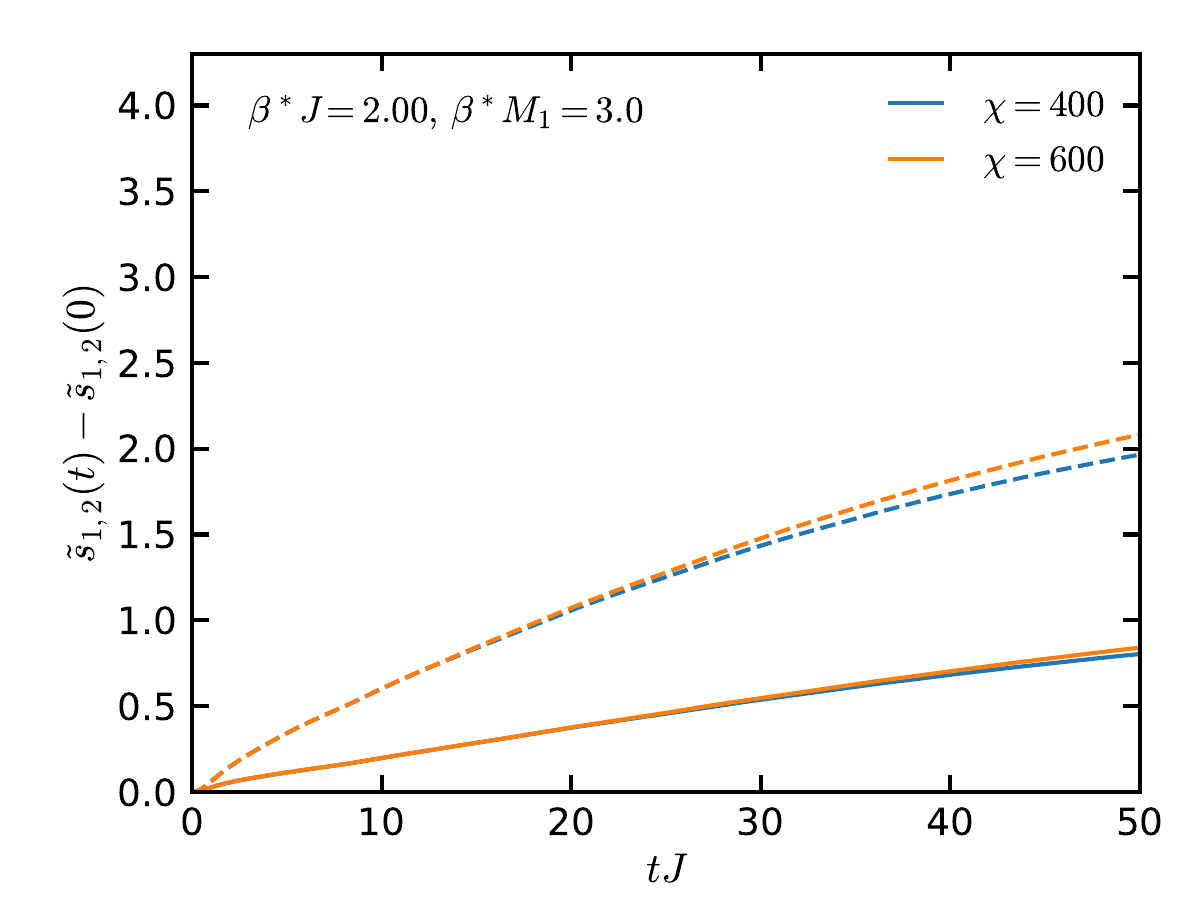}\\ 
   \circled{1}
\end{minipage}
\begin{minipage}{0.32\textwidth}
   \includegraphics[width=\textwidth]{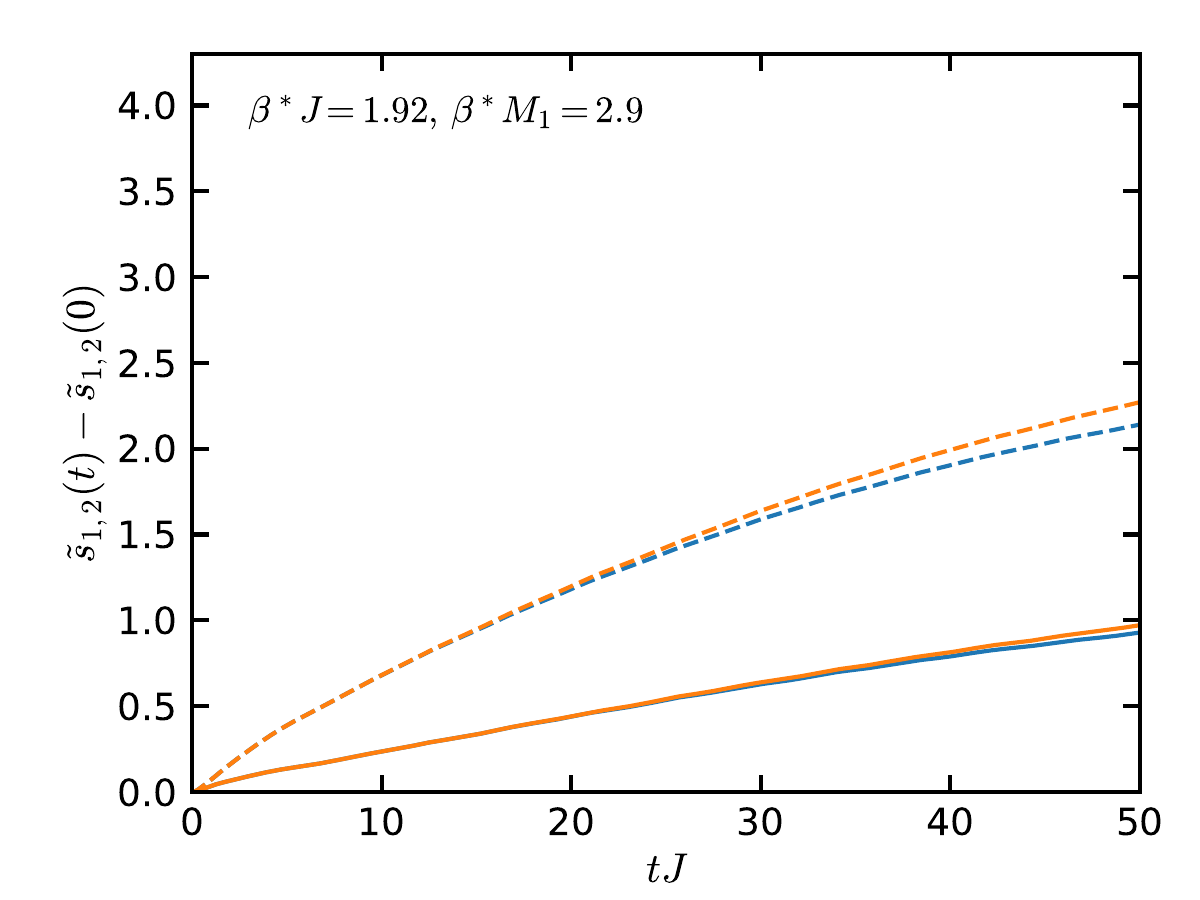}\\ 
   \circled{2}
\end{minipage}
\begin{minipage}{0.32\textwidth}
   \includegraphics[width=\textwidth]{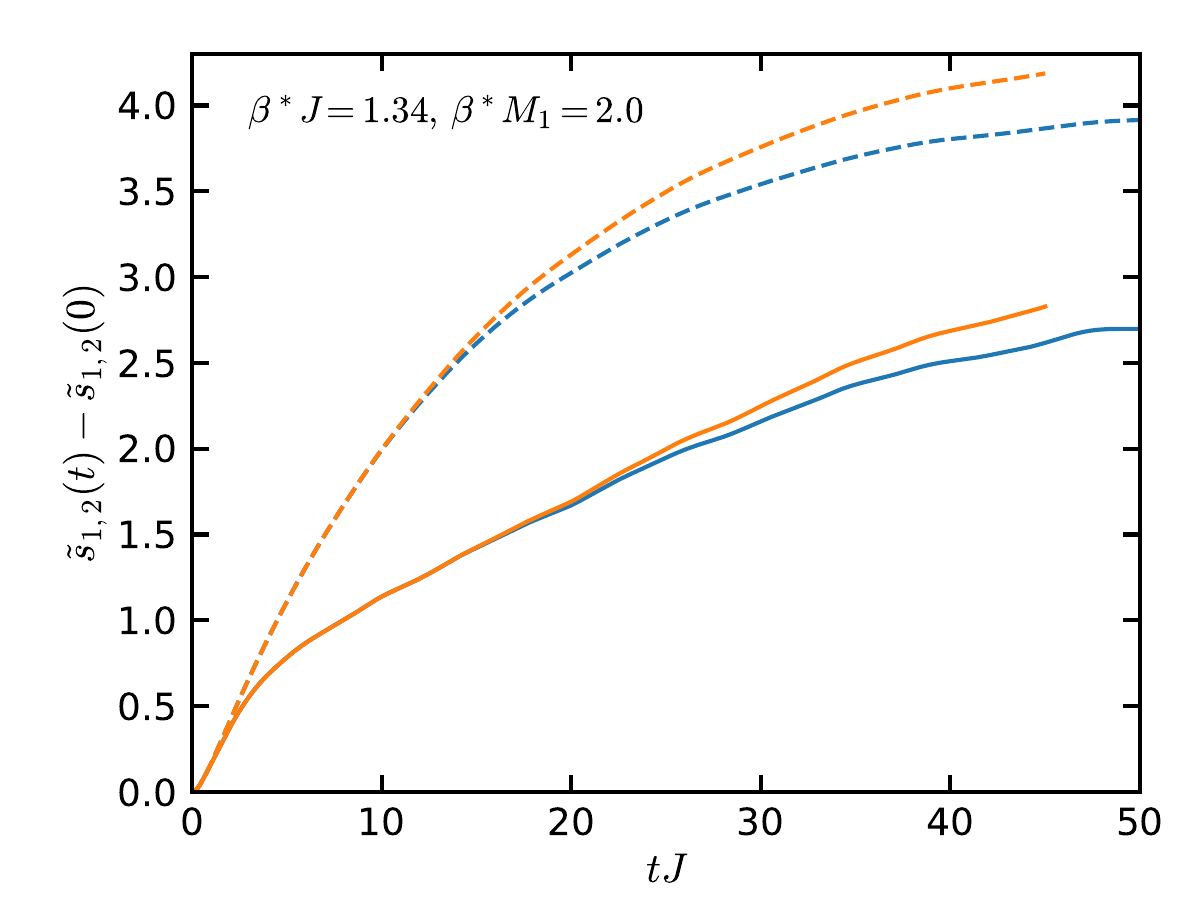}\\ 
   \circled{3}
\end{minipage}
    \caption{Time dependence of the reflected entropies for the quench protocols (1)-(3) (from left to right) with an initial temperature $\beta J=2$ and effective temperatures $\beta^*M_1$ as shown. The first (dashed) and second (solid) reflected entropies are shown for several bond dimensions (blue vs.\ orange curves), indicating that the results have not yet converged at late times.
    }
    \label{fig:s12_betaJ2_types}
\end{figure*}

\begin{figure*}[t]
    \centering
\begin{minipage}{0.32\textwidth}
   \includegraphics[width=\textwidth]{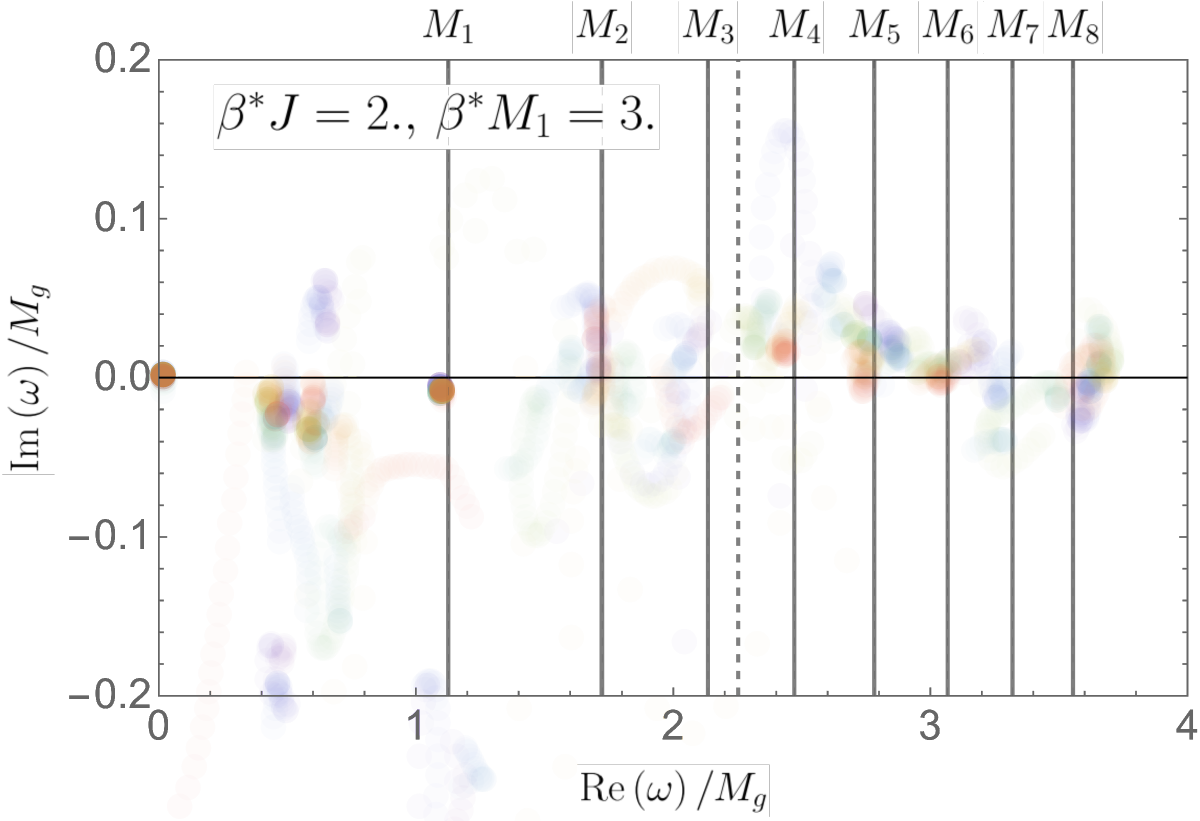} 
\end{minipage}
\begin{minipage}{0.32\textwidth}
   \includegraphics[width=\textwidth]{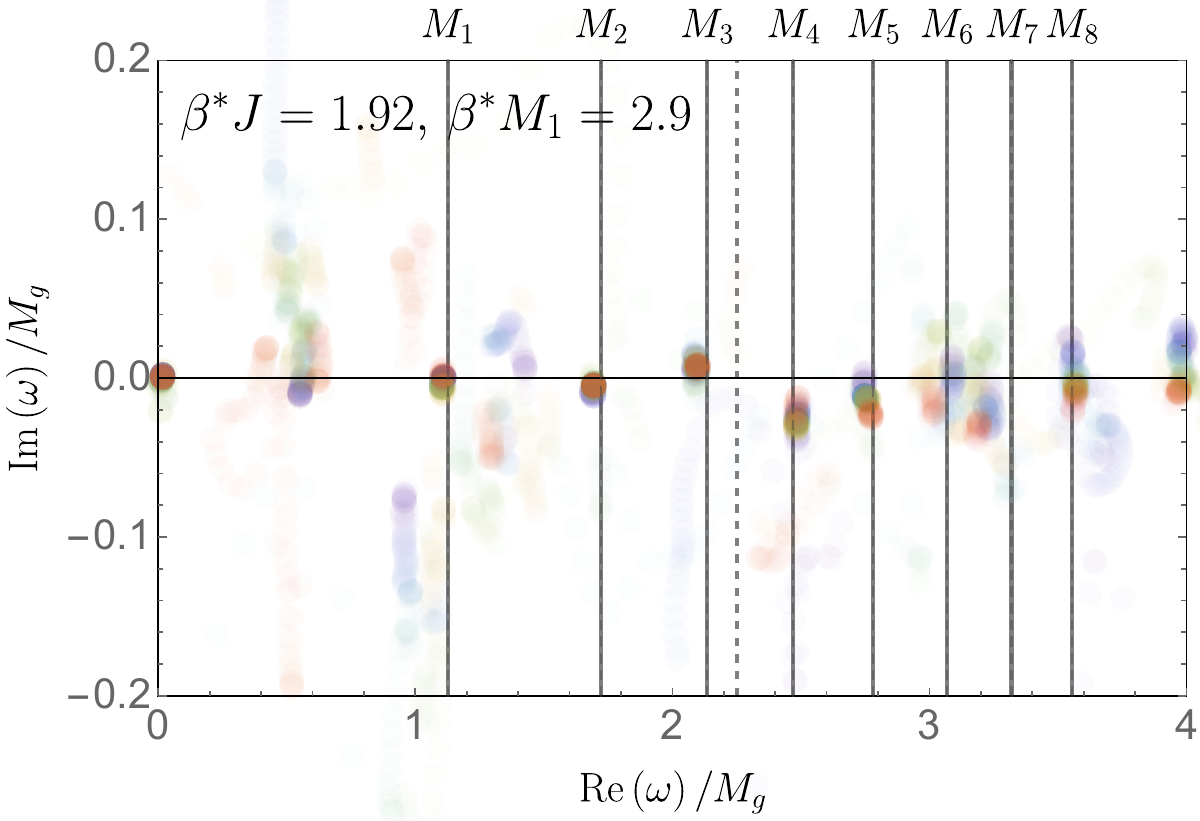} 
\end{minipage}
\begin{minipage}{0.32\textwidth}
   \includegraphics[width=\textwidth]{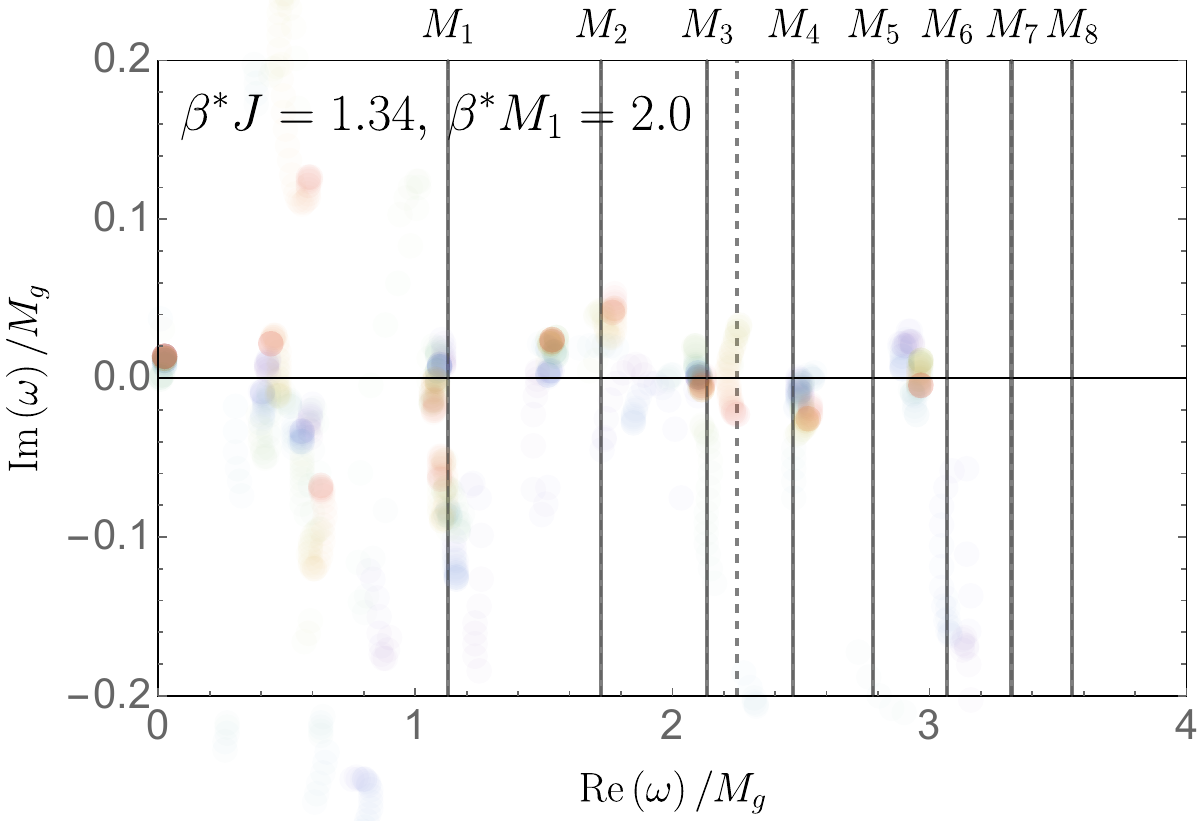} 
\end{minipage}
\begin{minipage}{0.32\textwidth}
   \includegraphics[width=\textwidth]{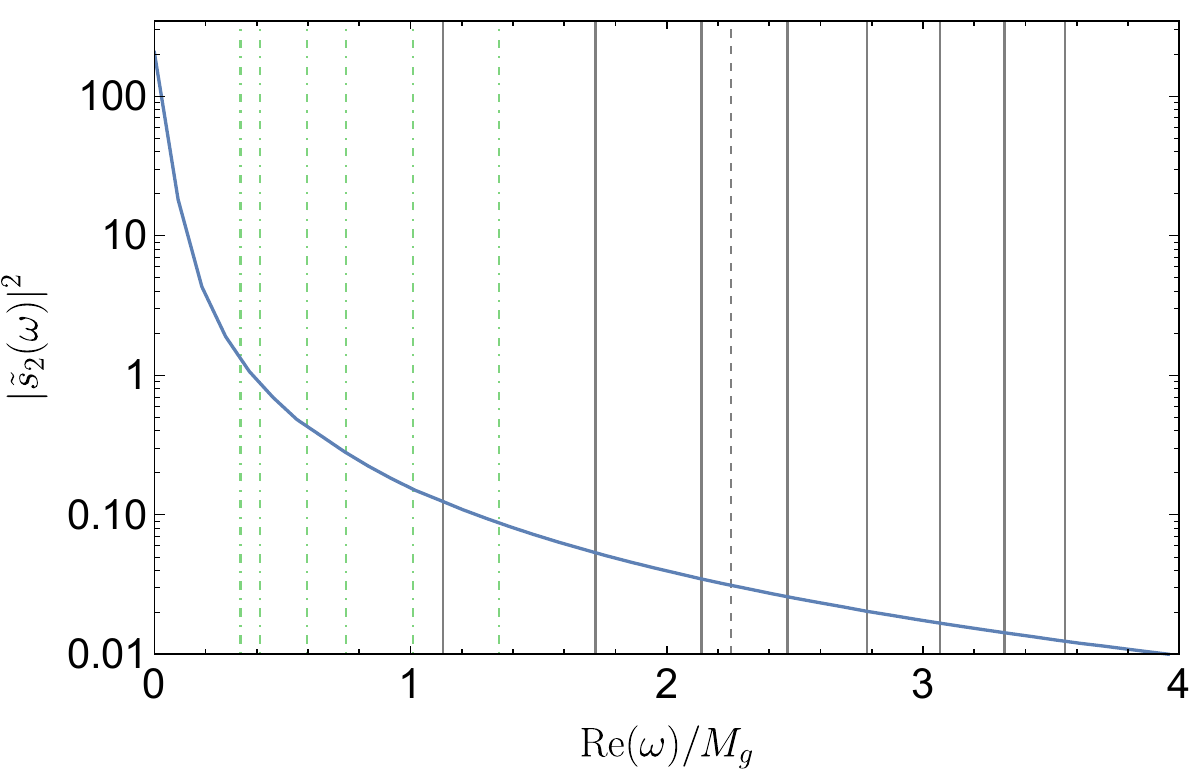}\\ 
   \circled{1}
\end{minipage}
\begin{minipage}{0.32\textwidth}
   \includegraphics[width=\textwidth]{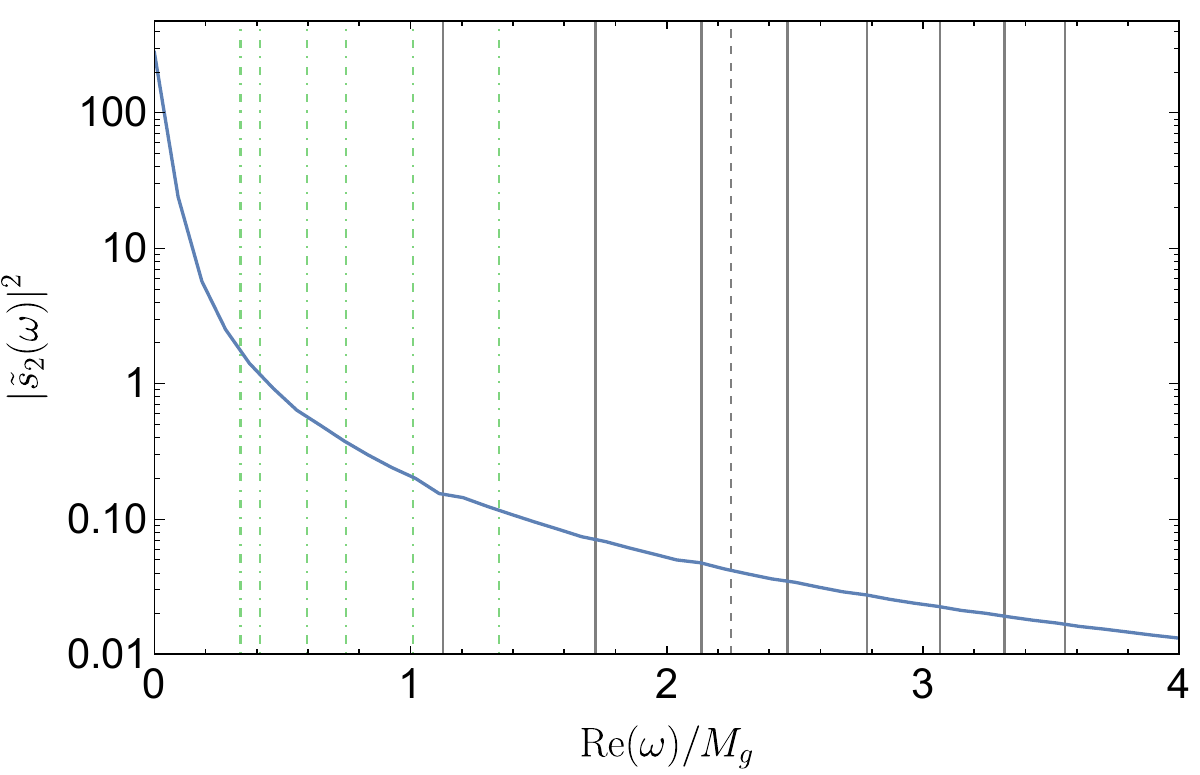}\\ 
   \circled{2}
\end{minipage}
\begin{minipage}{0.32\textwidth}
   \includegraphics[width=\textwidth]{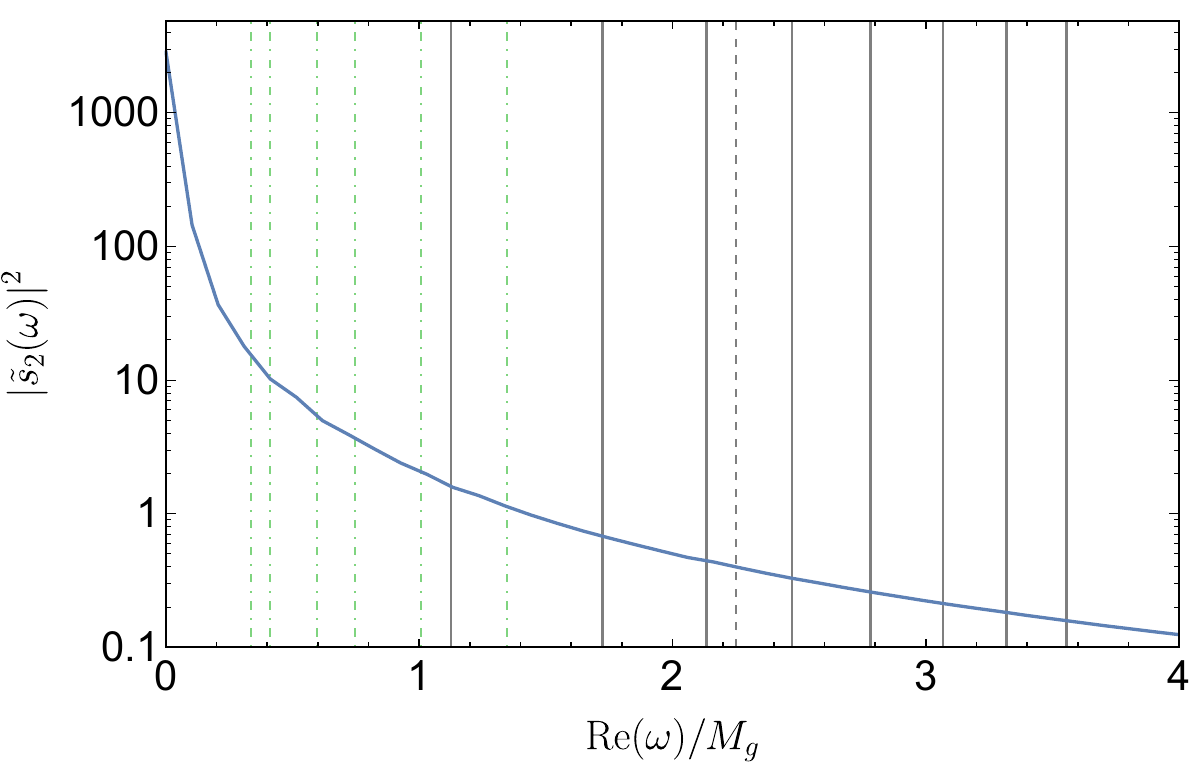}\\ 
   \circled{3}
\end{minipage}
    \caption{Results of the Prony signal analysis (top) and Fourier spectra (bottom) of $\tilde s_2$ for the quench protocols (1)-(3) (from left to right) with an initial temperature $\beta J=2$ and effective temperatures $\beta^*M_1$ as shown. Results are shown for the $\chi=600$ curves in Fig.\,\ref{fig:s12_betaJ2_types}. Background lines are as in Fig.\,\ref{fig:s2_FT_and_Prony_beta16}. The plots exemplify the effect of increased effective temperature through raising the initial temperature as compared to Fig.\,\ref{fig:s2_FT_and_Prony_beta16}: Except for the E$_8$ initial regime, meson states do not leave a considerable impact on the entanglement growth through entanglement oscillations.
    }
    \label{fig:s2_FT_and_Prony_beta2}
\end{figure*}

We now study the situation of raised effective temperature by starting from an initial thermal state at a higher temperature $\beta J=2$. In this case, the entanglement growth is larger and it becomes important to monitor the behavior of the iTEBD simulation in dependence of the bond dimension. The obtained reflected entropies are shown in Fig.\,\ref{fig:s12_betaJ2_types} for the quench protocols \circled{1}-\circled{3} (from left to right) for two different bond dimensions, $\chi$=400 (blue) and 600 (orange). The effective temperatures for protocol \circled{1} and \circled{2} are very close, $\beta^*M_1 \approx 3.0$ and $\beta^*M_1 \approx 2.9$, resulting in similar curves for the reflected entropies. For quench type \circled{3}, the effective temperature $\beta^*M_1 \approx 2.0$ causes an entanglement growth about twice as fast. At long times we observe that the results for both bond dimensions differ, indicating that they are not fully converged. This effect is milder for $\tilde s_2$ (both on an absolute and relative scale), which is most likely because $\tilde s_2$ is more local than $\tilde s_1$. Due to the larger entanglement growth, the discrepancy is larger for quench type \circled{3}.

Being aware of these limitations, we compare in Fig.\,\ref{fig:s2_FT_and_Prony_beta2} the Prony results and Fourier spectra of $\tilde s_2$ for the different quench protocols (for the highest bond dimension) for an overall qualitative assessment. In all examples, features in the complex frequency become fuzzier in the Prony analysis when compared to the quench results in Fig.\,\ref{fig:s2_FT_and_Prony_beta16} at lower temperatures. Some of the meson poles fully disappear, in particular for type \circled{1} and \circled{3} (top left and right panel). Observe also that the branch cuts indicating the continuum threshold disappeared in all types. The corresponding Fourier spectra are increased by several orders of magnitude and now flattened out in all cases. Although types \circled{1} and \circled{2} have a comparable effective temperature, the Prony result for the latter (top middle panel) still allows a much clearer identification of some meson states in the Prony analysis. This difference seems to be rooted in the fact that initial states were prepared in distinct phases with different properties.

\begin{figure}[t]
    \centering
   \includegraphics[width=0.49\textwidth]{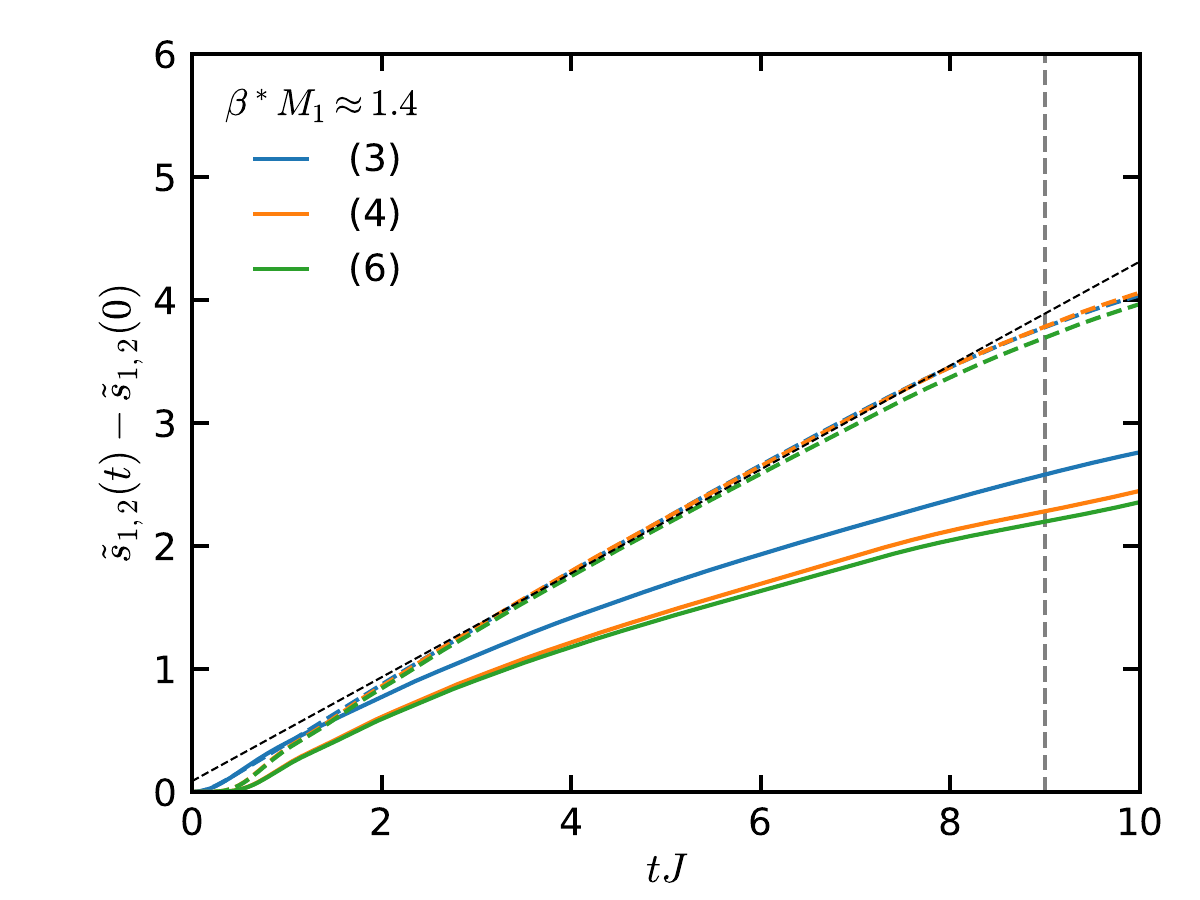} 
    \caption{Time dependence of the first (dashed) and second (solid) reflected entropy for quench protocols (3), (4) and (6) with an effective temperature $\beta^*M_1\approx1.4$. The vertical dashed gray line indicates convergence in the bond dimensions up to $tJ \lesssim 9$.
    The curves differ only mildly although initialized in different phases and temperatures, implying that the described features are induced by increasing the effective temperature.}
    \label{fig:s12_betaJ16_type4_and_typecomp}
\end{figure}

\begin{figure}[t]
    \centering
   \includegraphics[width=0.49\textwidth]{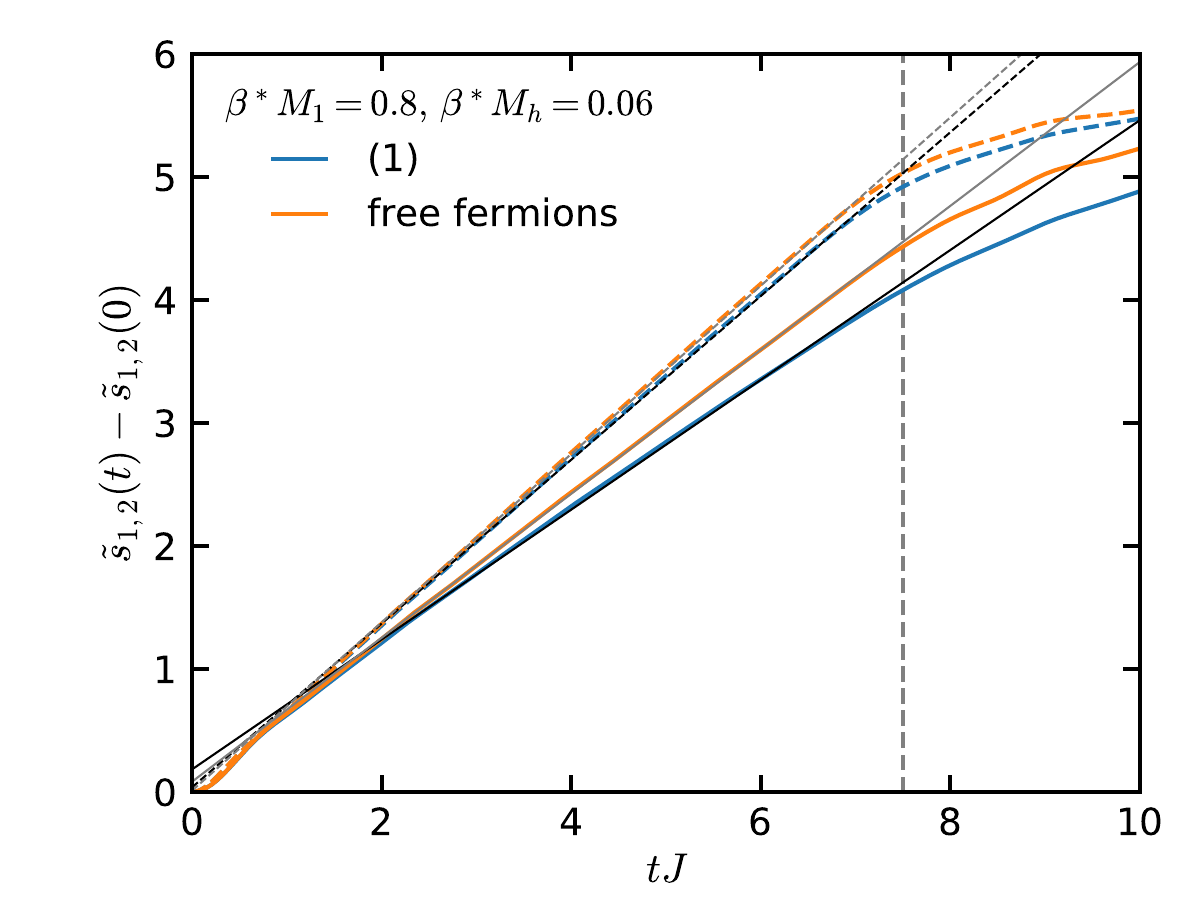} 
    \caption{Time dependence of $\tilde s_1$ (dashed blue) and $\tilde s_2$ (solid blue) for quench protocol (1) with an effective temperature $\beta^*M_1 \approx 0.8$. The vertical dashed gray line indicates convergence in the bond dimensions up to $tJ \lesssim 7.5$. The results for $\tilde s_{1,2}$ in the nonintegrable ferromagnetic phase (blue curves) are compared to a massive free fermion counterpart (orange curves) at the same effective temperature in units of the lattice spacing. The black and gray dashed and solid lines mark linear fit functions for $\tilde s_1$ and $\tilde s_2$ at early times, respectively.
    The results demonstrate the emergence of an overall linear growth behavior at high effective temperatures in the nonintegrable ferromagnetic phase, which is consistent with the interpretation of melted meson states which do not dominate the entanglement growth anymore. This behavior is analogous to the free fermion regime, in which no meson bound states are present.}
    \label{fig:s12_betaJ05_type1_and_ff}
\end{figure}

The overall phenomenological picture that arises from these analyses is that an increase of the effective temperature leads to an enhancement of the entanglement growth, which suppresses the impact of entanglement oscillations caused by meson states in the thermal bath. In absence of mesons, the quasiparticle model predicts a fully linear growth of entanglement entropies. We therefore aim in the following discussion to identify a linear growth of the reflected entropies as a signature of meson melting at the highest attainable temperatures. Leaning on the discussions in \cite{Scopa:2021gcx,Birnkammer:2022giy}, this phenomenological description can be alternatively understood in terms of meson densities.\,\footnote{We thank Alvise Bastianello for discussions on this point.} While low temperature quenches excite dilute mesons at rest, mesons start to become denser, scatter and move at finite velocity as the effective temperature increases. This superimposes a large entanglement growth to the entanglement oscillations. At very high temperatures, mesons are very dense and cannot be resolved individually anymore, since entanglement is spread very fast and entanglement oscillations are suppressed. This causes a linear entropy growth, analogous to the free fermion regime, which we interpret as meson melting. In this regime, mesons are not the relevant degrees of freedom for the system dynamics any longer.

In this line of reasoning, we consider in Fig.\,\ref{fig:s12_betaJ16_type4_and_typecomp} the time evolution of reflected entropies for the quench types \circled{3}, \circled{4} and \circled{6} with an (nearly) identical effective temperature $\beta^* M_1 \approx 1.4$. Due to the large entanglement growth, numerical results are converged in bond dimension only up to times $tJ \lesssim 9$ (shown by the gray dashed line). At later times the entropies saturate (cf.\ also the red curves in Fig.\,\ref{fig:s12_overview} for type \circled{4}), a signature of the truncation error. Altogether, there are only marginal differences for $\tilde s_1$ between all quench protocols and a small scaling difference for $\tilde s_2$ in type \circled{3} (blue solid curve) compared to the others. Since the initial states are in very different regimes (classical, ferromagnetic and critical), we interpret the overall agreement as a confirmation that the discussed effects on the entanglement growth are indeed thermally induced. Fig.\,\ref{fig:s12_betaJ16_type4_and_typecomp} therefore exhibits directly also a linear function, which is obtained by fitting $\tilde s_1$ in the time interval $2 \le t J \le 9$ as $0.09 + 0.42 (tJ)$. The data for $\tilde s_1$ are bounded by such a linear growth, but do not yet follow exactly this scaling behavior. On the other hand, the curves for $\tilde s_2$ grow less strongly in a nonlinear fashion. In this case, the time ranges for which the data are converged are too short to enable a robust Prony or Fourier signal analysis.

The highest effective temperature $\beta^*M_1 \approx 0.8$ was achieved for quench protocol \circled{1}. The results, displayed by blue curves (dashed for $\tilde s_1$, solid for $\tilde s_2$) in Fig.\,\ref{fig:s12_betaJ05_type1_and_ff}, converged in bond dimension up to $tJ \approx 7.5$ (indicated by the vertical dashed gray line). One can observe a strong growth of both reflected entropies. In the time interval $1.5 \le tJ \le 7.5$, we can fit the functions $0.04 + 0.67(tJ)$ and $0.18 + 0.53(tJ)$ to $\tilde s_1$ and $\tilde s_2$, respectively. In contrast to the simulations at intermediate temperatures, both reflected entropies now seem to follow this linear growth behavior after a brief initial time period with irregular growth induced through the quench itself. The same phenomenological picture arises independently in quench type \circled{2} (not shown here), i.e.\ with an initial state in the E$_8$ regime, at the same temperature. As alluded before, the existence of mesons necessarily implies the appearance of entanglement oscillations in the time evolution of entropic quantities, consistent with a quasiparticle model interpretation, if applicable in the nonintegrable range. Since in the present case we have $\beta^* M_1 < 1$ in the physical high-temperature regime, we explain the linear behavior through the fact that the mesons are melted, i.e.\ do not contribute entanglement oscillations to the overall growth of entanglement entropies anymore.

This interpretation is corroborated through a quench study in the (massive) free fermion regime. Complementary to protocol \circled{1}, we choose for that case a transverse quench using the parameters $(h_0,g_0,\beta)=(0.93,0,0.5) \rightarrow (h_1,g_1,\beta^*)=(0.9375,0,0.50)$. The resulting reflected entropies (orange curves) are compared in Fig.~\ref{fig:s12_betaJ05_type1_and_ff} to the previous nonintegrable counterpart (blue curves). Although the effective temperatures are not directly comparable anymore due to the different mass scales, the differences in the entanglement growth between these two quench types are relatively small. In particular, the free fermion data confirm the linear growth of the reflected entropies after the initial quench period. We can respectively fit the functions $0.01 + 0.68(tJ)$ and $0.08 + 0.59(tJ)$ to $\tilde s_1$ and $\tilde s_2$, which are shown as gray curves and can be compared to the black ones for the previous nonintegrable regime. These results confirm that a linear growth of reflected entropies is a property of the free fermion regime, which is in accordance with the quasiparticle model description. It therefore strengthens our interpretation on meson melting for the nonintegrable ferromagnetic case, since this qualitative behavior can be explained when meson states are not present in the physical system.

\subsection{Quench results in the high-temperature scaling limit}

\begin{figure*}[t]
    \centering
   \includegraphics[width=0.49\textwidth]{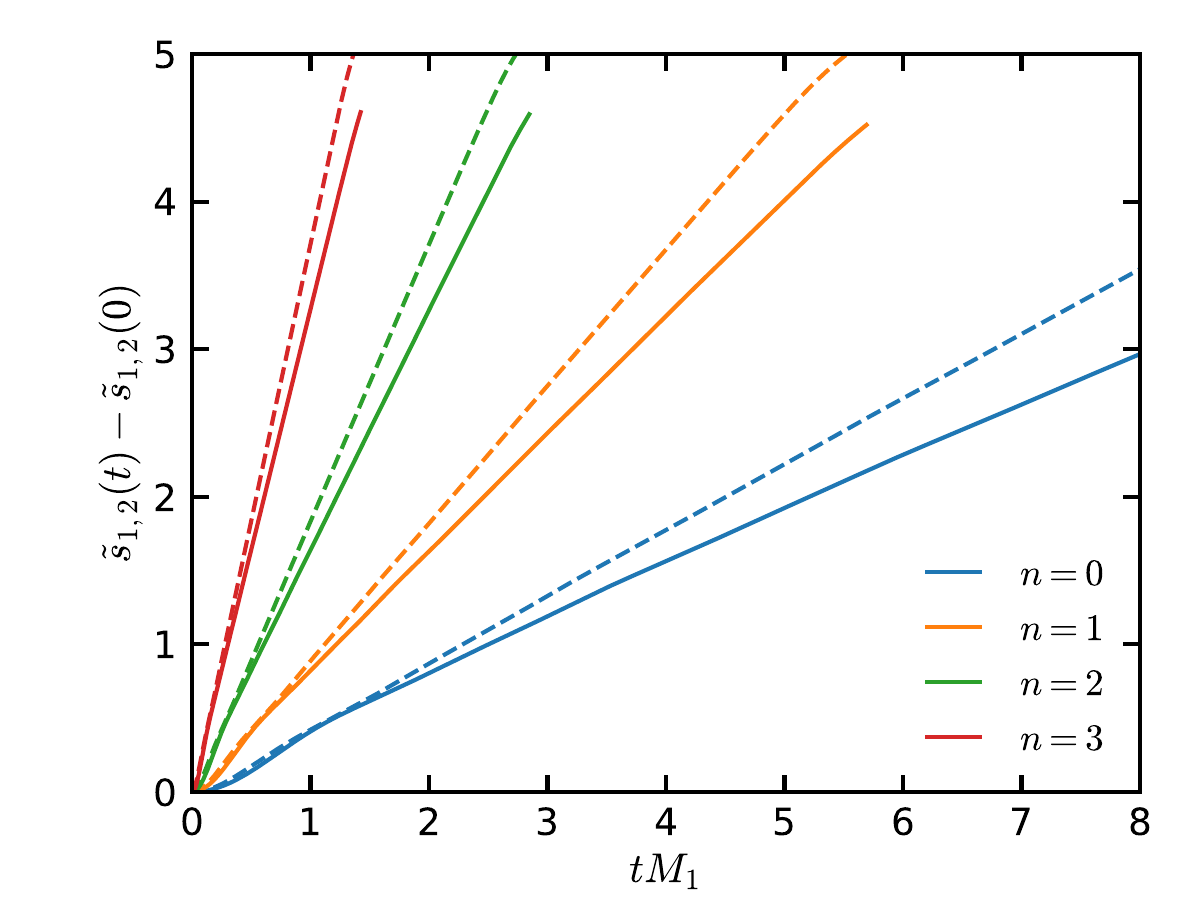} 
   \includegraphics[width=0.49\textwidth]{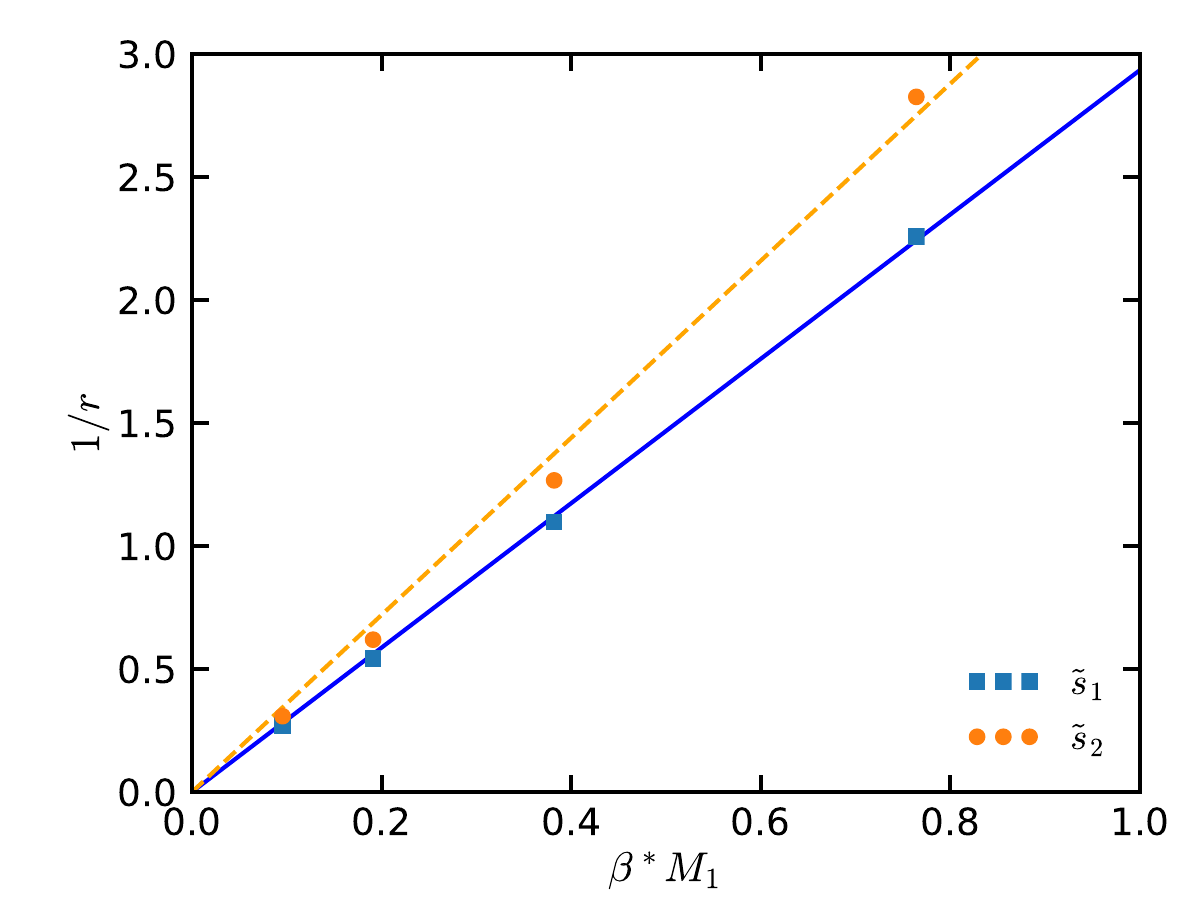} 
    \caption{Dependence of $\tilde s_1$ (dashed) and $\tilde s_2$ (solid) on the physical time $t M_1$ for quenches of type (1) in the high-temperature scaling limit, when mesons are expected to be melted (left panel). The post quench masses are varied according to the parametrization \eqref{eq:Mhg_params}, resulting in higher effective temperatures for increasing $n$. Only the time intervals of convergence are shown. The reflected entropies are growing more rapidly when approaching the scaling limit for increasing $n$ and can be fitted by linear functions of the form \eqref{eq:lin_s12}, whose inverse slopes $r^{-1}$ are plotted in the right panel in dependence on the inverse effective temperature $\beta^* M_1$ (blue squares for $\tilde s_1$, orange circles for $\tilde s_2$). The inverse slopes themselves approach the infinite temperature limit in a linear fashion (solid blue and dashed orange fit curves), implying a high-temperature dependence of the form \eqref{eq:s12_highT} for the reflected entropies.}
    \label{fig:contlimit1}
\end{figure*}

The results of the previous section have shown that we can extract meson states from entanglement oscillations of reflected entropies at low and intermediate temperatures in the nonintegrable ferromagnetic phase. Their masses match quantitative expectations of the Ising QFT. When the regime of high effective temperatures is reached, the linear entropy growth suppresses all oscillations. In this case, which we interpreted through the melting of mesons, the entanglement growth is very large and its QFT origin is not immediately visible anymore. Similarly to the simulations in thermal equilibrium, it then makes sense to consider the quench scenario also in the scaling limit in order to approach the QFT regime rigorously.

We therefore now focus on the scaling limit, defined as $M_h/J \to 0$ for fixed ratio $M_h/M_g$, for quench protocol \circled{1} at the highest effective temperatures. This setup leaves some degrees of freedom, since we can parametrize the pre-quench ($M_h^{pre}, M_g^{pre}$) and post-quench ($M_h^{post}, M_g^{post}$) masses as well as the initial $(\beta)$ and effective temperatures $(\beta^*)$. Here, we keep the ratios $M_h^{pre}/M_g^{pre} \approx 0.10$ and $M_h^{post}/M_g^{post} \approx 0.09$ constant as in the original quench setup, and vary the post-quench masses again according to the parametrization \eqref{eq:Mhg_params}. Since we perform a transverse quench, we have $M_g^{pre} = M_g^{post}$ and hence $M_h^{pre}/M_g^{pre} = M_h^{pre}/M_g^{post}$. 

In the first set of simulations, we choose, identically to the previous case, the initial temperature $\beta J = 0.5$, which results in different physical effective temperatures $\beta^* M_1$. The results for the reflected entropies are shown in the left panel of Fig.~\ref{fig:contlimit1}. The time dependencies of $\tilde s_1$ (dashed curves) and $\tilde s_2$ (solid curves) are shown in physical units, i.e.\ $t M_1$, where $M_1$ is the first meson mass read out from \cite{Fonseca:2006au}. We plot only the early time window, in which convergence is achieved. 
As we increase $n$, i.e.\ as we approach the critical point in the scaling limit $M_h/J \to 0$, the reflected entropies grow faster. In all cases, a linear growth behavior of the form
\begin{equation} \label{eq:lin_s12}
    \tilde s_{1,2}(t) = r\, t\, M_1 + s_{1,2}(t=0)
\end{equation}
can be deduced from the results after a very short initial quench phase. Here, $r$ denotes the slope of the linear dependence. The numerical fit results for $r^{-1}$ are shown in the right panel of Fig.~~\ref{fig:contlimit1} as a function of the (inverse) effective temperature $\beta^* M_1$. Both the results for $\tilde s_1$ (blue squares) and $\tilde s_2$ (orange circles) can be well fitted by linear functions through the origin as $r^{-1}_{\tilde s_1} = (2.93 \pm 0.02) \beta^*M_1$ $r^{-1}_{\tilde s_2} = (3.60 \pm 0.10) \beta^*M_1$. This, unsurprisingly, tells us that the slope diverges when reaching the infinite temperature limit. It, however, shows that this limit is approached with a proper scaling behavior, set only by the temperature. This means that the reflected entropies follow in the high-temperature regime, up to a numerical constant and shift, the time dependence
\begin{equation} \label{eq:s12_highT}
    \tilde s_{1,2}(t) \sim \frac{t}{\beta^*} .
\end{equation} 
This behavior is consistent with the interpretation that mesons are melted, since their mass scale does not leave an imprint on the time evolution anymore.

\begin{figure}[t]
    \centering
   \includegraphics[width=0.49\textwidth]{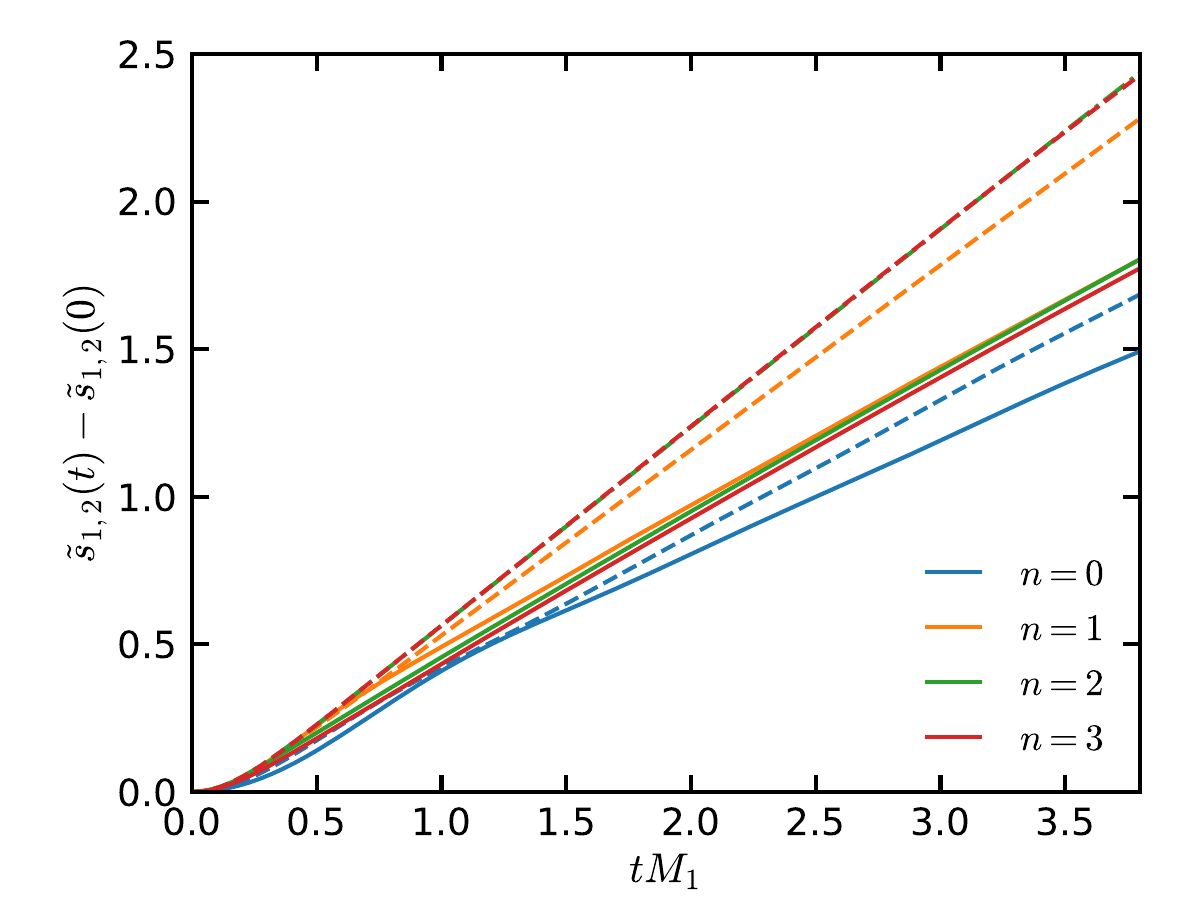} 
    \caption{Dependence of $\tilde s_1$ (dashed) and $\tilde s_2$ (solid) on the physical time $t M_1$ for quenches of type (1) in the high-temperature scaling limit. The post quench masses are varied according to the parametrization \eqref{eq:Mhg_params} at fixed physical temperature $\beta^* M_1 \approx 0.76$. The entropy curves converge for increasing $n$ and confirm an overall linear increase associated to mesons being melted.}
    \label{fig:contlimit2}
\end{figure}
As elaborated before, the scaling limit, in which the Ising QFT emerges, can be taken also in a different way. While we have so far considered a situation in which the effective temperature is raised, we now keep $\beta^* M_1 \approx 0.76$ fixed.\,\footnote{This parameter value corresponds to the blue $n=0$ curves in Fig.\,\ref{fig:contlimit1}.} Using the same parametrization \eqref{eq:Mhg_params} for the post-quench parameters, this implies the series $\beta^* J = \{0.5, 1, 2, 4\}$ for the effective temperatures in transverse quenches of type \circled{1}.\,\footnote{The initial temperature $\beta J$ follows from demanding \eqref{eq:betaeff}, and $M_h^{pre}$ is uniquely determined from the specified ratio $M_h^{pre}/M_g^{pre}$.} Fig.~\ref{fig:contlimit2} shows the resulting time dependencies of $\tilde s_1$ (dashed curves) and $\tilde s_2$ (solid curves). It becomes discernible that the first reflected entropy grows faster when $n$ is increasing from $n=0$ to $n=2$. The results for $\tilde s_1$ at $n=2$ (green dashed curve) and $n=3$ (red dashed curve) are coinciding, indicating that convergence is achieved. For $\tilde s_2$, convergence is approached slightly slower in the available time window (solid orange, green and red curves). These findings confirm the overall qualitative picture of a linear increase in absence of meson oscillations, which persists in the high-temperature scaling limit of the Ising QFT.

\section{Discussion and outlook}

In this article we initiated and performed an in-depth study of the meson melting phenomenon from a quantum information perspective. We consider the ($1+1$)-dimensional Ising QFT emerging in the continuum limit of the quantum Ising model as a test bench for a novel entanglement viewpoint on the topic. This is possible because the quantum Ising model contains nonperturbative meson bound states in integrable and nonintegrable ferromagnetic parameter regimes, which resemble the relevant confinement property with QCD. It therefore can provide important insights into phenomena manifested also by much more complicated theories. While this study is motivated by the analogous QCD effect, we however emphasize that in the present analysis we do not even attempt to draw any conclusions for QCD itself. 

From a modern QCD viewpoint, the meson melting mechanism is interpreted as a sequential process that strictly requires a dynamical treatment. While both QCD and holographic approaches indicate the melting of individual meson states through a thermal broadening of their corresponding peaks in the spectral function, in our previous work~\cite{Banuls:2019qrq} we could not observe an analogous feature in the complex frequency structure of retarded thermal equilibrium correlators in ferromagnetic parameter regimes of the ($1+1$)-dimensional Ising QFT. We therefore motivated in this paper a new paradigm to capture and describe meson melting characteristics via entropic quantities. For this purpose we employed TN simulations, which enable the approximation of thermal states as well as their subsequent real-time evolution. Moreover, TNs easily allow calculations of several entropies directly in the thermodynamic limit.

In particular, for a thermal equilibrium setting, we showed in section~\ref{sec:Renyi} that the scaling of the second R\'enyi entropy density is determined at low temperatures through an exponential damping (as a function of the inverse temperature $\beta$) set by the meson masses. In contrast, at high temperatures, the behavior is identical to that expected in a CFT (power-law dependence), meaning that meson states have been melted, i.e.\ the temperature sets the only scale in the physical system.

An independent, truly out of equilibrium probe is provided by the real-time evolution of reflected entropies after thermal quantum quenches (cf.\ section~\ref{sec:quench}). In this setting, we prepared initial thermal states in the ferromagnetic phase of the Ising model and implemented a quantum quench by simulating their evolution under a fixed post-quench Hamiltonian in the nonintegrable ferromagnetic phase. Whereas the initial state is not in equilibrium, we can identify an effective temperature corresponding to the energy density of the state. We could vary the effective temperature of this setup either through different parameter regimes in the pre-quench state or by modifying its initial temperature. While at low effective temperatures mesons give rise to entanglement oscillations (in line with the vacuum quenches in~\cite{kormos2017real}), we observed a linear entanglement growth at the highest achieved temperatures, meaning that the imprint of mesons is fully suppressed and that they are not the relevant degrees of freedom any longer. The latter observation is consistent with a quasiparticle model interpretation that excludes the presence of meson states, and with MPO simulations in the (massive) free fermion regime. We similarly infer from these dynamical properties that the meson melting process took place. For the particular theory under consideration, the ($1+1$)-dimensional Ising QFT, we do not see conclusive indications to decide whether this process is consistent with a sequential picture. Such a feature in principle could be discernible from the temperature dependent meson content in the Fourier spectra of reflected entropies. Both our static and dynamical results, however, suggest that the meson melting takes place as a smooth crossover, in analogy to QCD. In higher dimensions, there are in principle more geometric possibilities for the definition of a subsystem. From the general idea of the quasiparticle model we, however, expect that the same phenomenological picture of linear entanglement growth as a meson melting indication is valid also in this case.

Our analyses comprehensively show that R\'enyi or reflected entropies can serve as witnesses of the meson melting process both in static and dynamical settings at finite temperature. While we focused on the relativistic Ising QFT emerging in the IR limit of the corresponding spin chain model, it would be of course highly desirable to analyze a similar setting also in more complicated and higher-dimensional gauge theories. In view of the rapid progress in using both TN and quantum simulations for this purpose (see \cite{Banuls2019ropp,Banuls:2019bmf,Bass:2021bjv,Davoudi:2022cah} for an overview of current developments and future prospects for high-energy physics), one should remain optimistic about prospects of studying meson melting in these richer systems along the lines we developed in the present article. In particular, some of our considered entropic quantities, like R\'enyi entropies, became already measurable in quantum simulation experiments \cite{Brydges_2019}. This opens the opportunity to study similar physical systems and test our predictions beyond the classical computational realm. In a parallel vein, the recent work \cite{Rigobello:2021fxw} demonstrated the role of entanglement generation in ($1+1$)-dimensional QED scattering events using TNs. Following these lines of research, we are convinced that a quantum information based perspective can provide relevant insights into many open questions related to mesonic systems in particle and condensed matter physics.

\begin{acknowledgments}
We would like to thank Burkhard K\"ampfer and Rico Z\"ollner for useful discussions and correspondence. We thank Alvise Bastianello for comments on the manuscript and discussions on the topic. 
This work was partly supported by the Deutsche Forschungsgemeinschaft (DFG, German Research Foundation) under Germany's Excellence Strategy -- EXC-2111 -- 390814868, and by the EU-QUANTERA project TNiSQ (BA 6059/1-1). 
The work of JK was supported in part by a fellowship from the Studienstiftung des deutschen Volkes (German Academic Scholarship Foundation). 
JK is supported by the Israel Academy of Sciences and Humanities \& Council for Higher Education Excellence Fellowship Program for International Postdoctoral Researchers.
\end{acknowledgments}

\appendix

\section{Review of QCD and holographic approaches}
\label{app:review}

\subsection{Theoretical methods in QCD}
\label{app:review_methods}

The \textit{meson operator} entering formula~\eqref{eq:rhoImDR} of the spectral function is constructed as $M(\vec x,t) \equiv \bar{q}_f(\vec x,t)\,\Gamma\, q_f(\vec x,t)$ from the Dirac spinor $q_f$ of a quark with flavor $f$. Here, $\Gamma$ is the vertex operator - a combination of gamma matrices, which selects the spin and angular momentum. See, e.g., \cite{Rothkopf:2019ipj} for a detailed selection of possible choices. In the simplest cases, the identity $\mathds{1}$ or gamma matrix $\gamma^\mu$ can be chosen. 
Real-time properties such as strength and form of field fluctuations are captured by the time-ordered two-point function $D(\vec{x},t,\vec{x}_0,t_0) \equiv \langle \mathcal T M(\vec x,t) M^\dagger(\vec x_0,t_0) \rangle_\beta = \Tr\{\rho_\beta \mathcal T M(\vec x,t) M^\dagger(\vec x_0,t_0)\}$, where $\mathcal T$ is the the time-ordering operator and $\rho_\beta$ the thermal density operator. This quantity can be expressed as a path integral over the Schwinger--Keldysh contour that contains both the initial conditions (encoded in thermal density matrix elements) and the quantum dynamics (enforced by the QCD action). The time-ordered two-point function can be decomposed as follows:
\begin{align}
    D(\vec{x},\vec{x}_0,t,t_0) &= \frac{1}{2}\langle\{M(\vec x,t), M^\dagger(\vec x_0,t_0)\}\rangle_\beta \nonumber\\
    &\phantom{=}+ \frac{1}{2}\operatorname{sign}(t-t_0)\langle[M(\vec x,t), M^\dagger(\vec x_0,t_0)]\rangle_\beta  \nonumber\\
    &\equiv F(\vec{x},t,\vec{x}_0,t_0) - \frac{i}{2}\operatorname{sign}(t-t_0)\rho(\vec{x},t,\vec{x}_0,t_0) .
\end{align}
Here, we defined the \textit{statistical function} $F$, which measures the population of states, and the \textit{spectral function} $\rho$. 
Meson peaks in $\rho$ depend on the momentum $\vec p$, following the (relativistic) dispersion relation of the QFT. Above the \textit{continuum threshold}, given by twice the mass of the lightest particle, unbound particle pair peaks can appear within a continuous background structure. The \textit{thermal width} of any such peak is proportional to the inverse lifetime of the corresponding (quasi)particle. Another important quantity in this context is the \textit{binding energy} of a bound state, which is given by the difference of its mass peak to the continuum threshold.

QCD based methods for the calculations of spectral functions have intrinsic limitations, as we will concisely review here following the detailed discussion in \cite{Rothkopf:2019ipj}. 
First, effective field theory approaches make use of the fact that meson flavors of large mass can be integrated out to get a nonrelativistic approximation of QCD. The resulting expansion in energy scales is, however, truncated, limiting the physical content of the description. 
Furthermore, lattice QCD allows a nonperturbative treatment on a discretized spacetime lattice based on Monte Carlo sampling. However, the computations in taking the continuum limit are numerically extremely expensive. Both these methods give access to the Euclidean correlator $D_E(\vec x,\tau) = \langle M(\vec x,-i\tau) M^\dagger(\vec 0,0) \rangle_\beta$, which is related in Fourier space to the spectral function via an integral transformation. The extraction of $\rho(\vec p,\omega)$ from this relation for only a limited number of discrete, uncertainty-prone lattice data points limits the quantitative robustness of the results. 
An alternative approach is based on a potential method. To define it, consider the real-time Wilson loop
\begin{equation}
    W_\Box(r,t) = \mathcal P \exp\left[ i g \oint_{r\times t} A^\mu \mathrm dz_\mu \right]
\end{equation}
along a rectangular path of size $r$ times $t$ in the spatial and time direction. The static interquark potential $V_s$ is then defined as
\begin{equation} \label{eq:V_s}
    V_s(r) = \lim_{t\to\infty} \frac{i\partial_t W_\Box(r,t)}{W_\Box(r,t)} .
\end{equation}
This potential together with a mass and kinetic term defines a Hamiltonian, which governs the time evolution of the two-point meson correlator via the Schr\"odinger equation. The imaginary part of its solution in Fourier space yields the spectral function. The accuracy of this method is limited through the omittance of finite velocity corrections.

Overall, the presented methods of determining in-medium spectral functions for quark bound states at finite temperature provide the phenomenological picture of the meson melting process discussed in section~\ref{sec:prelim_pheno} in the main text.

\subsection{Insights from holographic models}
\label{app:review_holo}

Holographic approaches based on the gauge/gravity duality provide a framework to study high-temperature properties of quarkonium and to describe the melting of mesons. The underlying principle is to study a gravitational theory in a hyperbolic AdS spacetime with an additional holographic coordinate as the dual to a QFT in Minkowski spacetime. Since this duality translates the strong coupling regime of the QFT into a weakly coupled gravity theory, the considered problem can become computationally more tractable. Thermodynamic and entanglement properties of gravitational solutions admitting black hole horizons can then be identified with those of the gauge theory. (For general introductions into this broad field and further applications we refer to, e.g., \cite{Papantonopoulos:2011zz,Ammon:2015wua,Nastase:2015wjb,CasalderreySolana:2011us}.)
We briefly want to point out some major lessons regarding meson melting from these explorations, based on the recent series of papers \cite{Zollner:2016cgc,Zollner:2016rik,Zollner:2017nnh,Zollner:2017fkm,Zollner:2017ggh,Zollner:2018uep,Zollner:2020cxb,Zollner:2020nnt,Zollner:2021stb}, in which this topic was thoroughly discussed. 

The basic ingredient is the description of meson states as test particles in a fixed asymptotic AdS spacetime.\,\footnote{As test particles, different meson flavors do not react back between themselves nor with the metric and the dilaton.} The underlying action
\begin{equation}
    S = \int \mathrm d^4x \mathrm dz\, \sqrt{-g} \e^{-\phi} G(\phi) F_{mn}F^{mn}
\end{equation}
contains a dilaton field $\phi$, a flavor coupling function $G(\phi)$, and a $U(1)$ gauge field $V_m$ in the definition $F_{mn}=\partial_m V_n-\partial_n V_m$ of the field strength tensor. The dilaton breaks conformal symmetry and models the running coupling in QCD. Upon a field redefinition $\psi$ in Fourier space, the equation of motion of the gauge field takes in tortoise coordinates\,\footnote{The tortoise coordinate $\xi$ follows from the blackening function $f$ and the holographic coordinate $z$ as $\diff\xi=\diff z/f$. The mass $m$ is the zero component $p_0$ of the four-momentum.} the Schr\"odinger form
\begin{equation}
    \left( \partial_\xi^2 - [U_T-m^2] \right) \psi = 0 
\end{equation}
with a temperature-dependent potential $U_T$. Normalizable solutions of this equation yield a discrete mass spectrum $m_n^2$ ($n\in\mathds N$). The original \textit{soft-wall model} of \cite{Karch:2006pv} uses a quadratic ansatz (w.r.t.\ $z$) for the dilaton field and the simplest choice $G\equiv 1$ to enforce a linear spectrum of the form $m_n^2 = c_0 + c_1 n$, which matches experimental data of radial meson excitation spectra at $T=0$. At finite temperature, the potential well in $U_T(z)$ is decreasing, implying the existence of lesser bound states. In other words, the model is capable of reproducing the sequential melting of meson states. Further studies in \cite{Ghoroku:2005kg,Colangelo:2007pt,Colangelo:2008us,Fujita:2009wc,Colangelo:2009ra,Sui:2009xe,Cui:2011ag,Colangelo:2012jy} on refinements of this model found, however, unrealistic melting temperatures much below the QCD deconfinement scale $\mathcal O(\unit[150]{MeV})$. The authors of \cite{Zollner:2016cgc,Zollner:2016rik,Zollner:2017nnh,Zollner:2017fkm,Zollner:2017ggh,Zollner:2018uep,Zollner:2020cxb,Zollner:2020nnt} resolved that problem by finding one-parameter extensions of the metric and dilaton as well as a construction principle for the blackening function that result in consistent melting temperatures and, at the same time, exhibit thermodynamic properties consistent with the equation of state from lattice QCD. While this approach is based on ad-hoc ans\"atze for the background, the latter can be determined also self-consistently, for example, in a Einstein--dilaton model, defined by the action 
\begin{equation}
    S = \frac{1}{16\pi G_{\text{N}}^{(5)}} \int \mathrm d^4x \mathrm dz\, \left( R-\frac{1}{2}(\partial\phi)^2 - V(\phi) \right) ,
\end{equation}
where the dilaton potential $V(\phi)$ governs the thermodynamic system properties. It was found that then only non-trivial flavor functions $G(\phi)$ allow a consistent description of both meson trajectories (governed by $U_T$) and thermodynamic functions (as a crossover deconfinement transition). Based on a linear response framework for holographic settings, one can calculate the spectral function according to eq.~\eqref{eq:rhoImDR} from the boundary asymptotics of the bulk field. In~\cite{Zollner:2020nnt}, this led to the observation that the temperature, at which a peak forms in the spectrum might be much larger than the melting temperature following from the Schr\"odinger potential. 

Overall, one can fairly conclude that holographic methods do not yet reach the level to provide robust quantitative predictions of all QCD relevant features (meson trajectories, thermodynamic quantities, spectral functions). As noted also in \cite{Rothkopf:2019ipj}, some of the holographic models, e.g.\ in the recent work \cite{MartinContreras:2021bis}, predict the melting of spectral functions towards larger frequencies at high temperatures, which is in direct contradiction to QCD results.

In contrast to QCD based methods, holographic approaches provide, however, the advantage of giving access to entanglement measures. Starting with the work \cite{Klebanov:2007ws}, it was observed that holographic entanglement entropy, which according to the seminal work \cite{Ryu:2006bv,Ryu:2006ef} is given as a minimal bulk surface area, can serve as a probe (i.e.\ order parameter) of the confinement-deconfinement transition. Further explorations of this quantity for different holographic models mimicking QCD properties can be found in \cite{Lewkowycz:2012mw,Kim:2013ysa,Kol:2014nqa,Ghodrati:2015rta,Zhang:2016rcm,Dudal:2016joz,Knaute:2017lll,Ali-Akbari:2017vtb,Dudal:2018ztm,Li:2020pgn,Arefeva:2020uec,Lezgi:2020bkc,Ghodrati:2021ozc}. However, no such studies so far analyzed this phenomenon explicitly related to the meson melting process. Our TN simulations in this article therefore provide the first genuine study of entanglement properties in this physical context.

\section{Entanglement structure and tensor network ans\"atze}
\label{app:TNS}

The starting point of our considerations is a generic quantum many-body system with $N$ elementary constituents of local physical dimension $d$ living on discrete lattice sites. An arbitrary pure state vector $\ket{\Psi}$ in the tensor product Hilbert space $\mathcal H = \mathcal H_1 \otimes \mathcal H_1 \otimes \ldots \otimes \mathcal H_N$ of such a system can be written as
\begin{equation} \label{eq:Psi}
\ket{\Psi} = \sum_{i_1,i_2,\ldots,i_N=1}^d \psi_{i_1,i_2,\ldots,i_N} \ket{i_1}\otimes\ket{i_2}\otimes\cdots\otimes\ket{i_N} ,
\end{equation}
where $\psi_{i_1,i_2,\ldots,i_N}$ are complex coefficients w.r.t.\ some basis vectors $\ket{i_1},\ket{i_2},\ldots,\ket{i_N}$. The exponentially large set of $d^N$ wave function coefficients $\psi_{i_1,i_2,\ldots,i_N}$ easily exceeds classical computational resources for an exact representation, which is thus limited to very small systems. TN ans\"atze provide families of states with an efficient representation, in which the coefficients of the wave function are parametrized by a polynomial number of tensors (see  e.g.~\cite{Verstraete08,SCHOLLWOCK201196,Orus:2013kga,Okunishi2022review,Banuls2022review} for introductions to the topic).

In this work we make use of the matrix product state (MPS) and matrix product operator (MPO) ansatz. For a system with $N$ sites of physical dimension $d$, a MPO is an operator with the form
\begin{align}
\label{eq:MPO}
    O = \sum_{\substack{i_1,i_2,\ldots,i_N=1\\j_1,j_2,\ldots,j_N=1}}^d \Tr & ( A_{i_1 j_1}^{(1)} 
    A_{i_2 j_2}^{(2)} \cdots A_{i_N j_N}^{(N)}  )  \\
&    \ket{i_1}\bra{j_1}\otimes\ket{i_2}\bra{j_2}\otimes\cdots\otimes\ket{i_N}\bra{j_N}.  \nonumber
\end{align}
Here, the local rank-4 tensors $A_{\alpha\beta i_k j_k}^{(k)}$ $(k=1,\ldots,N)$ have two physical indices $i_k, j_k$ (corresponding to the bra and ket operation) and virtual indices $\alpha, \beta$ with bond dimension $\chi$. For open boundary conditions, the edge tensors ($k=1,\,N$) reduce to rank-3 tensors, instead. For fixed physical indices, the coefficient is a product of matrices, hence the nomenclature. The ansatz \eqref{eq:MPO} can be used to represent density operators, but the positivity condition cannot be imposed directly at the level of the individual tensors. It is however possible to further restrict the form of the tensors to enforce positivity, by using a purification. In particular, thermal states of local Hamiltonians admit an efficient approximation of this form, as discussed in section \ref{sec:methods}.

\section{Details of the transfer operator}
\label{app:transferop}

The MPO approximation of the unit cell of a thermal state can be represented graphically as a TN diagram as follows:
\begin{equation} \label{eq:rho_beta_MPO}
    \vcenter{\hbox{\includegraphics[width=0.38\columnwidth]{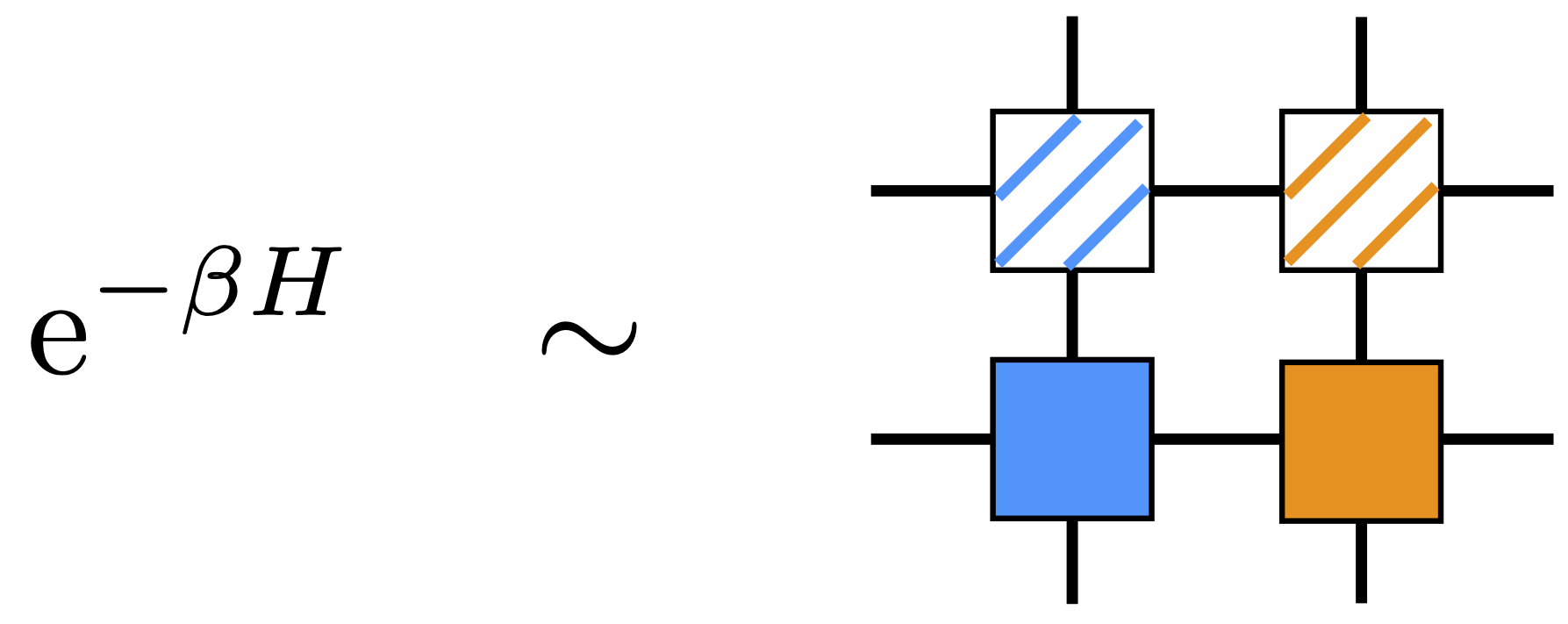}}}.
\end{equation}
Here, the two columns (blue and orange) indicate the repeating sites in the 2-site unit cell of the translational invariant chain. Each box represents a tensor $A_{\alpha\beta i_k j_k}^{(k)}$ as in the MPO definition \eqref{eq:MPO}. Virtual bond indices are drawn horizontally, physical ones vertically. Solid boxes symbolize $\sqrt{\rho_\beta}$ and striped boxes respectively the adjoint $\sqrt{\rho_\beta}^\dagger$. As discussed before, this construction ensures positivity of the state. The thermal transfer operator is then defined as 
\begin{equation} \label{eq:E_rho2}
    \vcenter{\hbox{\includegraphics[width=0.36\columnwidth]{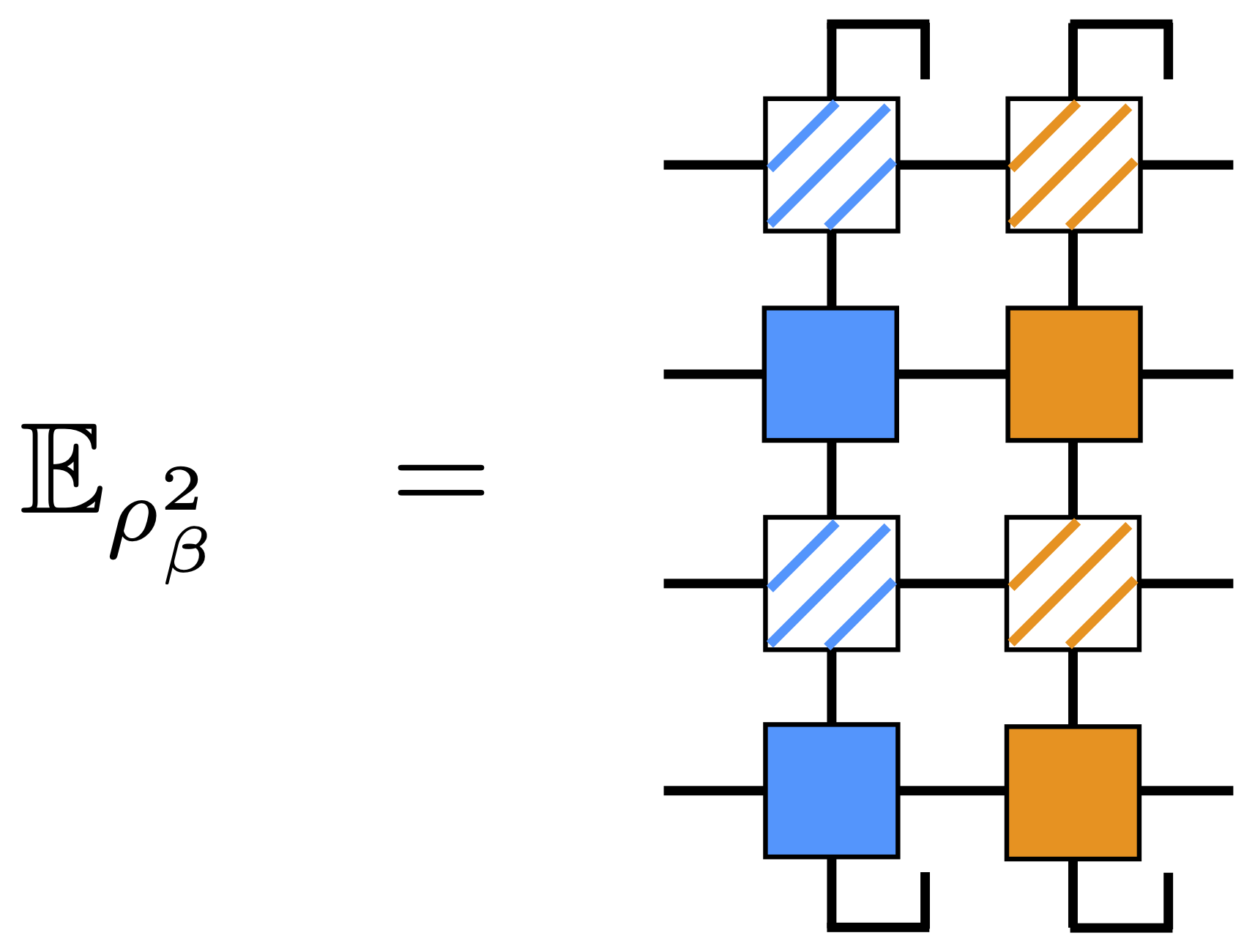}}},
\end{equation} 
where an overall trace over the physical indices is implied.

\section{Prony signal analysis}
\label{app:prony}

The basic idea of the Prony method \cite{peter2014generalized} is to reconstruct a function $f(t)$ as a sum of complex exponentials, 
\begin{equation} \label{eq:prony}
    f(t) = \sum_{k=1}^K c_k \e^{-i \, \omega_k\, t} ,
\end{equation}
where $c_k$ and $\omega_k$ are respectively complex coefficients and frequencies, and $K$ is the (variable) total number of modes. Prony based methods make use of the linearity of the ansatz \eqref{eq:prony} to determine $c_k$ and $\omega_k$ independently. In contrast to standard Fourier methods, this provides the advantage of extracting also complex valued frequencies. Observe in particular that real frequencies correspond to oscillations, while negative (positive) imaginary contributions give rise to an exponential decay (growth). In our previous work \cite{Banuls:2019qrq} we developed a signal analysis technique, which performs a Prony analysis on a finite subrange of the (time) domain of the variable $t$ and subsequently shifts the analysis window towards later times. Each identified mode is then visualized as in Fig.\,\ref{fig:s2_lowT_scaling} with a unique color in the complex frequency plane (on a scale from blue to red). Discrete frequencies will appear as stable modes across different analysis windows, whereas streaks (a sequence of poles in different time windows) indicate branch cuts.

\bibliographystyle{jk_ref_layout_wTitle} 
\bibliography{literature}

\end{document}